\documentclass[a4paper,11pt]{article}
\usepackage{jheppub} 
\usepackage[T1]{fontenc} 
\usepackage[utf8]{inputenc}
\usepackage{cancel, multirow, amsmath, dsfont, slashed, xcolor, fancybox}
\usepackage{soul}
\usepackage{cleveref}
\usepackage{booktabs}

\newcommand{\es}[2] {\begin{equation} \label{#1} \begin{split} #2 \end{split} \end{equation}}
\newcommand{\mpl}{M_{\rm pl}}
\newcommand{\D}{{\rm d}}

\newcommand{\f}{\mathfrak{f}}

\newcommand{\Dneff}{\Delta N_{\rm eff}}
\newcommand{\trh}{T_{\rm RH}}
\newcommand{\fa}{f_a}

\newcommand{\ba}[1]{\begin{align}\begin{split}#1\end{split}\end{align}}
\newcommand{\aqcd}[0]{a_{\rm QCD}}

\newcommand{\ch}[0]{c_H}

\newcommand{\cl}[0]{{\bf c}_L}

\newcommand{\cw}[0]{c_W}

\newcommand{\Hd}[0]{H^\dagger}

\newcommand{\GG}{G\widetilde{G}}
\newcommand{\WW}{W\widetilde{W}}
\newcommand{\BB}{B\widetilde{B}}

\newcommand{\Psibar}[0]{\overline{\Psi}}
\newcommand{\lfn}[0]{\Lambda_{\rm FN}}
\newcommand{\yh}{\mathcal{Y}_H}

\newcommand{\N}{{\cal N}}

\newcommand{\vckm}{V_{\rm CKM}}
\newcommand{\vckmd}{V_{\rm CKM}^\dagger}
\newcommand{\vecp}[0]{\boldsymbol{p}}
\newcommand{\veck}[0]{\boldsymbol{k}}

\newcommand{\journal}[1]{#1}

\title{Freezing-in the Axiverse}
\begin{document}
\author[a]{Christopher Dessert,}
\affiliation[a]{Center for Computational Astrophysics, Flatiron Institute, New York, NY 10010, USA}
\author[b]{Soubhik Kumar,}
\affiliation[b]{Institute of Cosmology, Department of Physics and Astronomy, Tufts University, Medford, MA 02155, USA}
\author[c]{and Joshua T. Ruderman}
\affiliation[c]{Center for Cosmology and Particle Physics, Department of Physics
New York University, New York, NY 10003, USA}
\emailAdd{cdessert@flatironinstitute.org}
\emailAdd{soubhik.kumar@tufts.edu}
\emailAdd{ruderman@nyu.edu}

\abstract{The presence of multiple light axions in the infrared is a generic feature of many ultraviolet (UV) scenarios. In many cases the number of axions ${\cal N}$ is ${\cal O}(10-100)$ or more.
Even in the scenario where these axions interact very weakly with the Standard Model (SM), the presence of ${\cal N}$ light axions poses a challenge to the stringent constraint on the number of relativistic degrees of freedom $N_{\rm eff}$. 
In order to remain agnostic about the UV, we adopt an effective field theory (EFT) approach, and parametrize the interactions of ${\cal N}$ axions with the SM to quantify the contribution to $N_{\rm eff}$.
We consider operators up to dimension six, uncovering one previously-unconsidered charge radius operator, and pay particular attention to the flavor structure of the axion-SM fermion couplings and consider EFTs based on anarchy, textures, and minimal flavor violation.
For various choices of such EFTs, we identify the discovery space for current and future cosmic microwave background surveys, including the Simons Observatory and CMB-HD\@.
We show this discovery space depends sensitively on the flavor structure and exhibits a rich interplay with terrestrial and astrophysical probes.}

\maketitle
\section{Introduction}

The Strong Charge-Parity (CP) problem in the Standard Model (SM) is the statement that the strong interactions appear to conserve CP symmetry while the weak interactions nearly maximally violate CP\@. Indeed, the Strong CP-violating parameter $\bar{\theta}$
is constrained to be $|\bar{\theta}|\lesssim 10^{-10}$~\cite{Abel:2020pzs}, 
while one would have expected $\bar{\theta}$ to be an ${\cal O}(1)$ number between $-\pi$ to $\pi$, since a vanishing $\bar{\theta}$ does not lead to any enhanced symmetry of the SM\@.
The existence of the quantum chromodynamics (QCD) axion is a hypothesized solution to this fine tuning problem. 
The QCD axion couples to gluons in the same way as how $\bar{\theta}$ appears in the SM Lagrangian.
The presence of the QCD axion leads to a dynamical relaxation of (the effective) $\bar{\theta}$ to a vanishing value as the QCD axion gets localized to the minimum of its potential, developed during the QCD phase transition~\cite{Peccei:1977hh,Peccei:1977ur,Weinberg:1977ma,Wilczek:1977pj}.
The QCD axion may also make up the dark matter (DM) of the Universe~\cite{Preskill:1982cy,Abbott:1982af,Dine:1982ah}. 

In minimal models, the QCD axion arises as the pseudo-Nambu-Goldstone boson of a global U(1) symmetry~\cite{Kim:1979if,Shifman:1979if,Dine:1981rt,Zhitnitsky:1980tq}.
In such cases, the QCD axion solution to the Strong CP problem may not be robust. 
Planck scale-suppressed higher dimensional operators can break the global $U(1)$ symmetry and modify the QCD axion potential significantly, altering the locations of the minima of the axion potential.
As a result, the QCD axion can get localized to a new minimum which does not correspond to a vanishing value of $\bar{\theta}$ and the Strong CP problem would reappear.
This is known as the `axion quality problem'~\cite{Georgi:1981pu,Dine:1986bg,Kamionkowski:1992mf,Holman:1992us,Barr:1992qq, Ghigna:1992iv}, since a robust solution to the Strong CP problem demands a `high quality' symmetry that is robust to various symmetry breaking effects. 

Models of the extra-dimensional QCD axion typically have such a high quality symmetry.
A classic example of this is the 5D $U(1)$ gauge field where the axion arises from the fifth component of the gauge field.
The higher dimensional gauge invariance~\cite{Choi:2003wr, Arkani-Hamed:2003xts}, or an electric one-form symmetry of the extra-dimensional theory~\cite{Reece:2025thc, Craig:2024dnl}, provides a more robust protection against symmetry breaking effects.
Contributions from symmetry breaking effects lead to exponentially-small corrections, in contrast to polynomial corrections found in minimal models.
This mechanism also plays out in string theory, where a large number $\N$ (as many as 491, found in Type IIB string theory analysis in~\cite{Demirtas:2018akl,Cicoli:2012sz}, \journal{or 181,200 in F-theory~\cite{Fallon:2025lvn}}) of pseudoscalars appear naturally as zero-modes of higher-dimensional gauge fields upon dimensional reduction.
This leads to the notion of an `Axiverse'~\cite{Witten:1984dg,Choi:1985je,Barr:1985hk,Svrcek:2006yi,Arvanitaki:2009fg} -- a universe with many axions. 

From a bottom-up perspective, the axion quality problem indicates that ultimately the axion solution to the strong CP problem is UV sensitive.  Whatever UV mechanism solves the axion quality problem may imply the existence of other light fields such as additional pseudo-Nambu-Goldstone bosons with approximate shift symmetries, such that the QCD axion is just one linear combination among a broader space of axion-like-particles (i.e., an axiverse). 
Hereon we will use ``axion'' to refer to any of the axion-like particles, and highlight the specific properties of the ``QCD axion'' when relevant.

The QCD axion is extremely light with mass $m_a \sim 6\,\mu{\rm eV}\, (f_a/10^{12}\,{\rm GeV})$~\cite{GrillidiCortona:2015jxo} where $f_a$ is the axion decay constant that controls the axion coupling with the SM, with astrophysical constraints requiring $f_a\gtrsim 10^9$~GeV~\cite{ParticleDataGroup:2024cfk}.
Interactions with the SM in the early universe may produce a ``cosmic axion background''~\cite{Turner:1986tb,Masso:2002np,Hannestad:2005df,Graf:2010tv,Salvio:2013iaa,Ferreira:2018vjj,Dror:2021nyr,DEramo:2021psx,DEramo:2021lgb,Dunsky:2022uoq,Langhoff:2022bij}, which can still be around as relativistic radiation today. 
The axion energy density $\rho_a$, in comparison with the photon energy density $\rho_\gamma$, can be parametrized by the $\Dneff$ parameter,
\es{}{
\Dneff \equiv {8\over 7}\left({11\over 4}\right)^{4/3} {\rho_a \over \rho_\gamma}\bigg\rvert_{\rm CMB},
}
evaluated during cosmic microwave background (CMB) decoupling.
For the QCD axion, the resulting $\Dneff$ is too small for current-generation CMB experiments. Indeed, the coupling
\es{}{
\mathcal{L} \supset c_G \dfrac{\alpha_s}{8\pi} \dfrac{a_{\rm QCD}}{f_a} \GG,
}
where $\alpha_s$ is the strong coupling parameter, leads to axion production via interaction with the thermal SM bath.
By dimensional analysis, this interaction rate goes as $T^3/f_a^2$.
Comparing this with the Hubble expansion rate, we obtain the temperature $T_d$ at which
the axion decouples from the SM bath, $T_d \sim 10^{11}$ GeV$(f_a/10^{12}\textrm{ GeV})^2$~\cite{Masso:2002np}.
If the reheat temperature of the Universe is above $T_d$, the QCD axion would obtain a thermal abundance.
In that case, the axion energy density today is that of an additional scalar degree of freedom diluted by all the SM entropy injection, contributing $\Delta N_{\rm eff} \approx 0.027$ (assuming standard cosmology).
This is below the sensitivity of current generation CMB surveys.
If the reheat temperature of the Universe is smaller than $T_d$, then the QCD axion does not thermalize and is only produced by `freeze-in'~\cite{Hall:2009bx} processes.
The resulting $\Dneff$ is even smaller in that case.

While the QCD axion alone contributes a small amount to $\Dneff$, one might expect the multitude of axions from the axiverse to give rise to a $\Dneff$ that is too large and ruled out by current the CMB and big-bang nucleosynthesis constraint, $\Delta N_{\rm eff} \lesssim 0.3$~\cite{Planck:2018vyg} at 95\% confidence.
That is, assuming each axion attains a thermal abundance, we have $\Delta N_{\rm eff} \approx 0.027\N$.
A scenario with $\N \sim {\cal O}(100)$ would then seem to be excluded by the observational bound, and only ${\cal N}\lesssim 10$ would be allowed.
However, the actual value $\Dneff$ depends on the SM couplings in a more subtle way.
To see this explicitly, consider $\N$ light axions $\phi_i$ with a coupling to QCD, so that
\es{}{
\mathcal{L} &= \dfrac{1}{2}\sum_{i=1}^{\N}(\partial_\mu \phi_i)(\partial^\mu \phi_i) + \sum_{i=1}^{\N} c_G^i \dfrac{\alpha_s}{8\pi} \dfrac{\phi_i}{f} \GG,
}
with some coefficients $c_G^i$.
Here we have assumed the axion masses are negligibly small, as will be quantified in Sec.~\ref{sec:ax_mass}.
The above form of the Lagrangian would still seem to suggest that all $\N$ axions would be produced by SM interactions, and each axion state will contribute to $\Dneff$. 
However, in the absence of masses, there is only the interaction `basis' with respect to which we can define the axion states. 
To identify this basis, we can perform an $SO(\N)$ field redefinition $a_i = \sum_i R_i^j \phi_j$ so that only one linear combination interacts with QCD:
\es{}{
\mathcal{L} =& \dfrac{1}{2}\sum_{i=1}^{\cal N}(\partial_\mu a_i)(\partial^\mu a_i) + c_G \dfrac{\alpha_s}{4\pi} \dfrac{a_{\rm QCD}}{f} \GG,
}
for $c_G = \sqrt{\sum_i(c_G^i)^2}$ and the QCD axion is identified as the linear combination $a_{\rm QCD} =(1/c_G)\sum_i c_G^i \phi_i$.
Thus, in this interaction basis, only the QCD axion interacts with the SM, while the remaining $({\cal N}-1)$ states are sterile.
As a result, we would again get $\Dneff \approx 0.027$, despite having $\N$ axions in the spectrum.
This exercise illustrates that in the presence of additional SM couplings in a scenario with ${\cal N}$ axions, a systematic degree of freedom counting and $\Dneff$ computation is necessary.
While Ref.~\cite{Gendler:2023kjt} discussed axion freeze-in \journal{(and decay~\cite{Yin:2025amn})} through their photon couplings for ${\cal N}>1$, the previous literature has primarily focused on the cosmological abundance of a single axion.
For example, Refs.~\cite{Higaki:2012ar, Cicoli:2012aq, Higaki:2013lra, Acharya:2015zfk, Gorbunov:2017ayg, Baer:2022fou} focused on axion production via moduli decay, while the freeze-in of individual axion-like particles \journal{both from SM couplings~\cite{Brust:2013ova, Baumann:2016wac, Arias-Aragon:2020shv, Green:2021hjh, DEramo:2021usm, Badziak:2024qjg, DEramo:2024jhn} and from SM+BSM couplings~\cite{Sakurai:2024cbi, DEramo:2023asj, DEramo:2023nzt} has been well-studied.}

In this work, we extend the above analysis to the presence of all additional operators mediating interactions between axions and the SM\@. 
We give the first systematic derivation of the EFT of the axiverse with $\N$ axions coupled to the SM scalar, fermion, and gauge sectors at operator dimension $d=5$ and $6$.
We also uncover a previously-unknown operator in this context, and show that $d\geq 7$ operators are not relevant for freeze-in. 
We find that each dimension-5 operator, of the form $(\partial_\mu \phi_i)\mathcal{O}^\mu_{\rm SM}$, couples a single axion state to the SM\@. 
On the other hand, dimension-6 operators, which schematically are of the form $(\partial \phi_i \partial \phi_j)\mathcal{O}_{\rm SM}$, generically couple the entire axion spectrum to the SM since each dimension-6 operator is quadratic in the axion field. 
We show that at dimension five, the number of physical axion degrees of freedom that interact with the SM is equal to the number ${\cal N}_{\rm ind}$ of independent dimension-5 operators in the EFT\@. Assuming similar strengths for all dimension-5 operators, and ${\cal N} > {\cal N}_{\rm ind}$, then: $\Dneff \propto {\cal N}_{\rm ind}$. This is the only known (to us) case where the number of independent operators of a certain dimension appears in a physical observable sensitive to beyond the Standard Model effects. At dimension six, instead, $\Dneff \propto {\cal N}$, the total number of states in the spectrum.

We use this EFT description to compute $\Dneff$ in the axiverse, assuming a standard radiation-dominated cosmology, with arbitrary reheating temperatures $\trh$ and axion decay constants $f_a$, and for a variety of assumptions about the origin of axion-SM couplings.
We place new constraints using recent Planck~\cite{Planck:2018vyg}, ACT~\cite{ACT:2025tim}, and SPT~\cite{SPT-3G:2025bzu} datasets, and forecast sensitivities for Simons Observatory~\cite{SimonsObservatory:2018koc,SimonsObservatory:2025wwn}, CMB-S4~\cite{CMB-S4:2016ple}, and CMB-HD~\cite{Sehgal:2019ewc}. 
We find that the flavor structure of the axion-fermion couplings is important in determining the observability of the cosmological axion background in the near future.
Adopting a bottom-up perspective we find that existing CMB data requires that the Universe reheated to a couple orders of magnitude below $f_a$, or hosts very few axions. 
Our result emphasizes the necessity of performing top-down computations of axion-SM couplings to sharpen the predictions for $\Dneff$.

The rest of this work is structured as follows.
In Sec.~\ref{sec:axEFT}, we first review the EFT coupling of SM fields with one axion, and then generalize that discussion to $\N$ axions. We take into account appropriate field redefinitions to remove redundant operators.
In Sec.~\ref{sec:benchmark}, we focus on several benchmark EFTs motivated by KSVZ~\cite{Kim:1979if,Shifman:1979if} and DFSZ~\cite{Dine:1981rt,Zhitnitsky:1980tq}-type couplings. We pay attention to the flavor structure by considering anarchy~\cite{Hall:1999sn}, Froggatt-Nielsen textures~\cite{Froggatt:1978nt}, and minimal flavor violation~\cite{DAmbrosio:2002vsn}.
In Sec.~\ref{sec:computation}, we compute the freeze-in production rate of the axions. We demonstrate that for dimension-5 axion couplings, the freeze-in abundance (at temperatures much larger than the weak scale) can be computed analytically, even including the full phase space distribution of the axions.
Finally, in Sec.~\ref{sec:result}, we use these rates to derive the results for $\Dneff$. We derive constraints from existing CMB data, highlight the interplay with astrophysical and flavor constraints, and identify targets for future CMB surveys.
We conclude in Sec.~\ref{sec:conc}.

\section{Axion Couplings in Effective Field Theory}\label{sec:axEFT}

\subsection{Single Axion EFT}

Before discussing an EFT containing $\N$ axions, it is useful to review the EFT of a single axion $\phi$ coupled to the unbroken phase of the SM~\cite{Bauer:2017ris,Bauer:2020jbp,Biekotter:2023mpd,Galda:2021hbr}. This EFT can most generally be written at dimension five as (in four-component Dirac fermion notation),
\es{eq:SingleAxion}{
\mathcal{L}^{(5)}_{\rm single} = \dfrac{\phi}{f}\left(c_G \dfrac{\alpha_s}{8\pi}  \GG + c_W \dfrac{\alpha_2}{8\pi}  \WW + c_B \dfrac{\alpha_1}{8\pi} \BB\right) + \sum_F \dfrac{\partial_\mu \phi}{f}{\Psibar}_F {\textbf c}_F\gamma^\mu\Psi_F + c_H \dfrac{\partial_\mu \phi}{f} H^\dag i\overset{\leftrightarrow}{D_\mu}H,
}
where $G$, $W$, and $B$ are, respectively, the strong, weak, and hypercharge field strengths, $H$ is the SM Higgs field, and $F = Q, u, d, L$, and $e$ runs over the SM chiral fermion multiplets.
In our convention, $Q$ and $L$ are left-handed, while $u, d$, and $e$ are right-handed Dirac fermions.
The $\textbf{c}_F$'s are $3\times 3$ complex Hermitian matrices depicting the flavor couplings of the axion.
The action is not invariant under axion-dependent chiral rotations proportional to hypercharge $Y$, baryon number $B$, and individual lepton number $L_n$, $n \in \{e,\mu,\tau\}$. In particular, these rotations act on the SM fields as
\es{eq:rot}{
	\Psi_F \longrightarrow& \exp\left(i\mathcal{Y}_F\theta_Y(x) + iB_F\theta_B(x) + i\sum_n L_F\theta_n(x)\right)\Psi_F, \\
    H \longrightarrow& \exp\left(i\mathcal{Y}_H\theta_Y(x)\right)H,
}
where $\mathcal{Y}_F$, $B_F$, and $L_F$ are the hypercharge, baryon, and lepton numbers of the multiplet $F$, and ${\cal Y}_H$ is the hypercharge of $H$.
We can use these five transformations, given by $\theta_Y, \theta_B, \theta_{e,\mu,\tau}$ to remove five redundant operators from~\eqref{eq:SingleAxion}. To see this, we compute the change of the action under the transformation~\eqref{eq:rot}; the result amounts to the replacements:
\es{}{
	\left(\ch \dfrac{\partial_\mu \phi}{f}\right)\Hd i\overset{\leftrightarrow}{D_\mu} H &\longrightarrow \left(\ch \dfrac{\partial_\mu \phi}{f} - \mathcal{Y}_H \partial_\mu \theta_Y\right)\Hd i\overset{\leftrightarrow}{D_\mu} H, \\
	\Psibar_F\left({\bf c}_F\dfrac{\cancel{\partial} \phi}{f}\right)\Psi_F &\longrightarrow \Psibar_F\left({\bf c}_F\dfrac{\cancel{\partial} \phi}{f} - \mathcal{Y}_F \cancel{\partial} \theta_Y \mathit{I} - B_F \cancel{\partial} \theta_B \mathit{I} - L_F {\rm diag}(\cancel{\partial} \theta_e,\cancel{\partial} \theta_\mu,\cancel{\partial} \theta_\tau)\right)\Psi_F, \\
    c_W \dfrac{\alpha_2}{8\pi} \dfrac{\phi}{f} \WW &\longrightarrow \dfrac{\alpha_2}{8\pi}\left(c_W \dfrac{\phi}{f} - 3\theta_B -  \sum_n \theta_n\right)\WW, \\
    c_B \dfrac{\alpha_1}{8\pi} \dfrac{\phi}{f} \BB  &\longrightarrow \dfrac{\alpha_1}{8\pi} \left(c_B \dfrac{\phi}{f} + 3 \theta_B +  \sum_n \theta_n \right)\BB.
}
The first two replacements can be derived by tracking the changes to the Higgs and fermion kinetic terms, while the last two replacements follow from anomalous rotations under $U(1)_B$ and $U(1)_{L_n}$.
Since the field redefinitions are not anomalous under $SU(3)_c$ and $U(1)_{B-L}$, $c_G$ and $c_W + c_B$, respectively, remain unchanged.
In addition, we also get two contributions at dimension six,
\es{eq:dim6_corr}{
\Delta {\cal L}^{(6)} = - 2c_H{\cal Y}_H \partial_\mu\theta_Y {\partial_\mu \phi \over f}|H|^2 + \yh^2 (\partial_\mu \theta_Y)^2 |H|^2,
}
from the dimension-5 axion-Higgs coupling and the Higgs kinetic term, respectively.

Various choices of rotation angles $\theta_Y, \theta_B, \theta_{e,\mu,\tau}$ can remove different terms from the Lagrangian. It is conventional to remove the axion-Higgs dimension-5 operator with the rotation $\theta_Y = 2c_H \phi/f$, which rotates into the flavor-conserving axion-fermion couplings.
This choice of $\theta_Y$ implies, based on Eq.~\eqref{eq:dim6_corr},
\es{}{
\Delta {\cal L}^{(6)} = - {c_H^2 \over f^2}(\partial_\mu\phi)^2 |H|^2.
}
There is no conventional choice to remove the remaining redundant operators, but without loss of generality, we will choose to set $c_W$ and the diagonal of $\cl$ to zero with the choices
\es{}{
\theta_n &= \left(\cl^{nn}+\ch\right)\dfrac{\phi}{f}, \\
\theta_B &= \left(\dfrac{1}{3}c_W - \dfrac{1}{3}{\rm Tr}(\cl) - c_H\right)\dfrac{\phi}{f}.
}
We see that the axion couples to $\underbrace{3}_{\text{gauge}}+\underbrace{1}_\text{Higgs}+\underbrace{5 \times 9}_\text{fermion}-\underbrace{5}_\text{redundant} = 44$ independent operators in the EFT at dimension five. Explicitly,
\es{eq:SingleFixedBasis}{
\mathcal{L}^{(5)}_{\rm minimal} =& c_G \dfrac{\alpha_s}{8\pi} \dfrac{\phi}{f} \GG + c_B \dfrac{\alpha_1}{8\pi} \dfrac{\phi}{f} \BB + \sum_{Q,u,d,e} \dfrac{\partial_\mu \phi}{f}{\Psibar}_F {\textbf c}_F\gamma^\mu\Psi_F + \sum_{n\neq m}\dfrac{\partial_\mu \phi}{f}{\Psibar}_L^n ({\textbf c}_L)_{nm}\gamma^\mu\Psi_L^m,
}
where we have redefined all transformed coefficients to maintain the original notation.
Note, if based on UV considerations, we have fewer than 44 couplings in the EFT to begin with, we cannot remove those couplings further in general, but rather only reshuffle among different operators. For example, if we only have axion coupling to the three gauge bosons, we cannot remove them entirely without them giving rise to additional (fermion) couplings.

\subsection{Axiverse EFT}

Now consider the more general effective field theory of $\N$ axions $\phi_i$ coupled to the SM, where $i$ runs from $1$ to $\N$:
\es{eq:GenLag}{
\mathcal{L} &= \dfrac{1}{2}\sum_i(\partial_\mu \phi_i)(\partial^\mu \phi_i) - \dfrac{1}{2}\sum_i m_i^2\phi_i^2 + \mathcal{L}_{\rm SM} + \mathcal{L}_{\rm int}^{(5)}, \\
\mathcal{L}_{\rm int}^{(5)} &=\, \sum_i c^i_G \dfrac{\alpha_s}{8\pi} \dfrac{\phi_i}{f} \GG + \sum_i c^i_2 \dfrac{\alpha_2}{8\pi} \dfrac{\phi_i}{f} \WW + \sum_i c^i_1 \dfrac{\alpha_1}{8\pi} \dfrac{\phi_i}{f} \BB  \\
& + \sum_{i,F} \dfrac{\partial_\mu \phi_i}{f}\Psibar_F {\textbf c}_F^i\gamma^\mu\Psi_F + \sum_i c_H^i \dfrac{\partial_\mu \phi_i}{f} H^\dag i\overset{\leftrightarrow}{D_\mu}H.
}
In App.~\ref{sec:general}, we provide further explanations that lead to this effective Lagrangian.
We assume that the bare masses $m_i$, that could come from some UV physics, are small enough such that the axions are effectively massless at the time of the CMB decoupling (see Sec.~\ref{sec:ax_mass} for more details).

\subsubsection{Dimension-5 EFT}

In Eq.~\eqref{eq:GenLag} we have $49 \times \N$ axion-SM operators, but as in the case with a single axion, not all of them are independent. Firstly, we note that at  dimension five all such operators are \textit{linear} in the axion fields. This means we can instead work with the linear combinations $a_\mathcal{O}$ that couple to a SM operator $\mathcal{O}$ as $a_\mathcal{O}\mathcal{O}$. For example, only the linear combination $a_\textrm{QCD} \propto \sum_i c^i_G \phi_i$ couples to the gluons and is defined as the {\it QCD axion}. However, since the couplings are independent in the absence of additional symmetries, the linear combination $a_{\mathcal{O}_1}$ is not in general related to any other $a_{\mathcal{O}_2}$, so one cannot construct a unique interaction basis.
A simple choice for the basis is to use the Gram-Schmidt 
orthogonalization procedure to write the interaction Lagrangian in upper triangular form (see App.~\ref{sec:general} for more details).
So schematically, we arrive at
\es{eq:axi_full}{
\mathcal{L} &= \dfrac{1}{2}\sum_i(\partial_\mu a_i)(\partial^\mu a_i) - \sum_{i,j}\dfrac{1}{2}a_i M_{ij}a_j + \mathcal{L}_{\rm SM} + {\cal L}^{(5)}_{\rm int}, \\
{\cal L}^{(5)}_{\rm int} &= c_G \dfrac{\alpha_s}{8\pi} \dfrac{a_{\rm QCD}}{f} \GG + [\textrm{couplings of } a_\textrm{QCD}\textrm{ to all other operators}] \\
& + c_W \dfrac{\alpha_2}{8\pi} \dfrac{a_W}{f} \WW + [\textrm{couplings of } a_W \textrm{ to all other operators \textit{except} QCD}] \\
& + c_B \dfrac{\alpha_1}{8\pi} \dfrac{a_B}{f} \BB + [\textrm{couplings of } a_B \textrm{ to all other operators \textit{except} QCD and SU(2)$_L$}] \\
& + \ldots \\
& + [\N - \N_{\rm ind}\ \textrm{sterile axions}],
}
where $\N_{\rm ind}$ is the number of independent axion-SM operators in Eq.~\eqref{eq:GenLag}.
Note, while the mass matrix $M_{ij}$ is not diagonal in this new basis, the effect of mass mixing will not be important for the light axions considered in this work (see Sec.~\ref{sec:ax_mass}, and Refs.~\cite{Kitajima:2014xla,Daido:2015bva,Daido:2015cba,Cyncynates:2021xzw,Cyncynates:2022wlq,Cyncynates:2023esj,Chadha-Day:2023wub,Dunsky:2025sgz,Murai:2023xjn,Murai:2024nsp,Murai:2025wbg} for scenarios where mass mixing of axions is relevant).
Therefore, we do not consider the mass matrix from now on.
For an explicit example of the above steps with ${\cal N}=3$ and ${\cal N}_{\rm ind}=2$, see App.~\ref{sec:example}.

We will also make the conservative assumption that we can drop the additional couplings in the square brackets of~\eqref{eq:axi_full}, or equivalently, that the interaction-basis axions are orthonormal: $a_i \cdot a_j = \delta_{ij}$. Considering those additional interactions would only enhance the production rate of axions and hence $\Dneff$, and therefore this approach gives a lower bound on the freeze-in yield. 
We then arrive at the following Lagrangian, where each SM operator interacts with an independent axion:
\es{eq:AxiverseLagrangian}{
\mathcal{L}^{(5)}_{\rm axiverse} &= c_G \dfrac{\alpha_s}{8\pi} \dfrac{a_{\rm QCD}}{f} \GG + c_W \dfrac{\alpha_2}{8\pi} \dfrac{a_W}{f} \WW + c_B \dfrac{\alpha_1}{8\pi} \dfrac{a_B}{f} \BB \\
& + \sum_F \Psibar_F^n c_{F,nm} \dfrac{\partial_\mu a_F^{nm}}{f}\gamma^\mu\Psi_F^m + c_H \dfrac{\partial_\mu a_H}{f} H^\dag i\overset{\leftrightarrow}{D_\mu}H.
}
Here $c_G a_{\rm QCD} = \sum_i c_G^i\phi_i$, $c_W a_W = \sum_i c_2^i \phi_i$, $\sum_i c_{F,nm}^i \phi_i = c_{F,nm}a_{F}^{nm}$, and so on.
We have also explicitly written out the generation indices $\{n,m\}\in\{1,2,3\}$ and defined $a_F^{nm}$ as the (possibly flavor-violating) axion which mediates a coupling between the $n$th and $m$th flavors of multiplet $F$. 

This manipulation makes evident that there are at most 49 relevant parameters rather than $49\N$.
However, we show there are still additional redundant operators that can be removed by appropriate field redefinitions, as with a single axion. Under the hypercharge, baryon, and lepton number rotations~\eqref{eq:rot} given by $\theta_Y$, $\theta_B$, $\theta_{n}$, we now have
\es{}{
	\left(\ch \dfrac{\partial_\mu a_H}{f}\right)\Hd i\overset{\leftrightarrow}{D_\mu} H &\longrightarrow \left(\ch \dfrac{\partial_\mu a_H}{f} - \mathcal{Y}_H \partial_\mu \theta_Y\right)\Hd i\overset{\leftrightarrow}{D_\mu} H, \\
	\Psibar_F\left({\bf c}_F\dfrac{\cancel{\partial} a_F}{f}\right)\Psi_F &\longrightarrow \Psibar_F\left({\bf c}_F\dfrac{\cancel{\partial} a_F}{f} - \mathcal{Y}_F \cancel{\partial} \theta_Y \mathit{I} - B_F \cancel{\partial} \theta_B \mathit{I} - L_F {\rm diag}(\cancel{\partial} \theta_e,\cancel{\partial} \theta_\mu,\cancel{\partial} \theta_\tau)\right)\Psi_F, \\
    c_W \dfrac{\alpha_2}{8\pi} \dfrac{a_W}{f} \WW &\longrightarrow \dfrac{\alpha_2}{8\pi}\left(c_W \dfrac{a_W}{f} - 3\theta_B - \sum_n \theta_n\right)\WW, \\
    c_B \dfrac{\alpha_1}{8\pi} \dfrac{a_B}{f} \BB  &\longrightarrow \dfrac{\alpha_1}{8\pi} \left(c_B \dfrac{a_B}{f} + 3 \theta_B + \sum_n \theta_n \right)\BB.
}
In particular, we can again rotate to the conventional basis by removing the axion-Higgs operator with the hypercharge rotation $\theta_Y = 2 c_H a_H/f$, setting $c_H\to0$.
Note that because the operators are linear in the axion, this is equivalent to removing the axion-Higgs coupling $c_H^i$ for every mass eigenstate $\phi_i$. This changes the axions $a_F$ that the fermions couple to, so that
\es{}{
\Psibar^n c_{F,nm} \dfrac{\partial_\mu a_F^{nm}}{f}\gamma^\mu\Psi^m \to \Psibar^n \left(c_{F,nm} \dfrac{\partial_\mu a_F^{nm}}{f} - 2c_H\mathcal{Y}_F\mathit{I} \dfrac{\partial_\mu a_H}{f}\right. \\
- B_F \partial_\mu \theta_B \mathit{I} - L_F {\rm diag}(\partial_\mu \theta_e,\partial_\mu \theta_\mu,\partial_\mu \theta_\tau)\bigg)\gamma^\mu\Psi^m.
}
We similarly make the combined $B$ and $L$ rotations to remove $\cw$ and the diagonal elements of $\cl$. Similar to the case of the single axion, we need
\es{}{\label{eq:GSrotations}
\theta_n &= c_{L,nn}\dfrac{a_L^{nn}}{f} + \ch \dfrac{a_H}{f}, \\
\theta_B &= \left(\dfrac{1}{3}c_W\dfrac{a_W}{f} - \dfrac{1}{3}{\rm Tr}\left(\cl\dfrac{a_F}{f}\right) - c_H \dfrac{a_H}{f}\right).
}
Note again that these transformations are simply that for a single axion repeated for each mass eigenstate. Removing these five operators leaves us with $44$ independent axions coupled to the SM, the same as for the single-axion case. This is because the axions are coupled linearly, which is not the case at dimension six. 

Finally, we arrive at the Lagrangian we will use to make predictions for $\Dneff$ (at dimension five), in which two axions are coupled to the gauge fields and 42 axions are coupled to fermions, given by
\es{eq:Lagrangian}{
\mathcal{L}^{(5)}_{\rm axiverse} =& \dfrac{\alpha_s}{8\pi} \dfrac{a_{\rm QCD}}{f_a} \GG + c_B \dfrac{\alpha_1}{8\pi} \dfrac{a_B}{f_a} \BB + \sum_{F,n,m \neq L,i,i} \Psibar_F^n c_{F,nm} \dfrac{\partial_\mu a_F^{nm}}{f_a}\gamma^\mu\Psi_F^m,
}
where we have redefined the $c$'s to account for the above field redefinitions and also defined $f_a = f/c_G$ as usual.
The condition on the fermion sum forbids axion coupling with the diagonal elements of ${\bf c}_L$. We emphasize that the removal of those couplings, and the axion coupling to $\WW$, is a basis choice. Other choices are possible, but will not affect physical observables.
The QCD coupling will give a mass to $a_{\rm QCD}$ upon confinement and we will track the cosmological abundance of $a_{\rm QCD}$ accordingly.

\subsubsection{Dimension-6 EFT}
\label{sec:dim-6}
There are two dimension-6 shift-symmetric SM-gauge-invariant operators for $\N$ axions
\es{eq:Dim6}{
\mathcal{L}_{\rm int}^{(6)} =& \sum_{i,j}\dfrac{c^{ij}_{aH}}{f_a^2}(\partial_\mu \phi_i)(\partial^\mu \phi_j)H^\dag H + \sum_{i,j}\dfrac{c^{ij}_{aB}}{f_a^2}(\partial_\mu \phi_i)(\partial_\nu \phi_j) B^{\mu\nu}.
}
These operators are fundamentally different from those at dimension five because they are quadratic in the axion fields. The first operator also mediates the lowest-order nonredundant coupling to the Higgs, and has previously been considered in~\cite{Bauer:2022rwf}.
The correction analogous to~\eqref{eq:dim6_corr}, originating from removing the dimension-5 Higgs coupling, can be absorbed into the coefficients $c_{aH}^{ij}$. Note that the Higgs coupling is symmetric in $i\leftrightarrow j$, while the hypercharge coupling is antisymmetric. The latter class of operator, known as a charge-radius operator, has been considered before~\cite{Bai:2012yq,Pierce:2014spa}, but not in the context of the axiverse.  Finally, replacing $B$ with $\tilde{B}$ results in an operator that vanishes upon using the Equations of Motion (EOM).

The operator with the Higgs coupling is written in an arbitrary basis.
In a basis where the kinetic term takes a canonical form, and the dimension-5 operators are in the Gram-Schmidt basis, the matrix $c_{aH}^{ij}$ is not diagonal in general.
However, a key feature of this interaction is that for every nonzero eigenvalue of $c_{aH}$, there is an associated axion which is coupled to the SM Higgs.
The number of such axions is equal to $\textrm{rank}(c_{aH})$. 
For a generic $c_{aH}$, we expect it to be full rank, so that at dimension six all $\N$ axions are coupled to the SM\@. 
This follows directly from the quadratic nature of the operator, which means that in the axion vector space it is characterized by a Wilson coefficient matrix rather than a vector.
In Sec.~\ref{sec:dim6Higgs} we compute the freeze-in abundance of axion production from the process $H H^\dagger \rightarrow \phi_i \phi_j$, neglecting the inverse process.
This is a good approximation except near the decoupling temperature of the dimension-6 operator.
Thus, the production of the $i$-th axion is proportional to 
\es{eq:lamah_i}{
\sum_{j\neq i} |c_{aH}^{ij}|^2 + 2|c_{aH}^{ii}|^2 \equiv |\lambda^{i}_{aH}|^2.
}
Therefore, being agnostic about the precise form of the $c_{aH}^{ij}$ matrix, we will just specify the vector $|\lambda^{i}_{aH}|$.

The latter operator, coupling to $B^{\mu\nu}$, is also quadratic but differs because $c_{aB}$ is skew-symmetric. It can be brought to a block-diagonal form where each axion kinetically mixes with only one other by an orthogonal transformation. However, the orthogonal matrix that implements this rotation in general does not commute with the matrices needed to perform the Gram-Schmidt orthogonalization at dimension five; therefore, while the rotation
\ba{
\sum_{ij}\dfrac{c^{ij}_{aB}}{f_a^2}(\partial_\mu \phi_i)(\partial_\nu \phi_j) B^{\mu\nu} \to \sum_i^{\lfloor \N/2 \rfloor}\dfrac{\lambda^{i}_{aB}}{f_a^2}(\partial_\mu a_{[i})(\partial_\nu a_{i+1]}) B^{\mu\nu}.
}
is possible, we do not perform it. Here $\lfloor x \rfloor$ is the floor function and the brackets in the indices denote anti-symmetry in $i \leftrightarrow i+1$. The eigenvalues of $c_{aB}$ are $\pm i\lambda^i_{aB}$ and, if $\N$ is odd, also zero.

Interactions such as $\phi_i+\phi_j\to B$, for on-shell $B$, have zero rate since $\partial_\mu B^{\mu\nu} \propto k^2 \epsilon^\nu + (\epsilon \cdot k)k^\nu=0$. However, we can use the EOM to rewrite, 
\ba{
(\partial_\nu \phi_{[i})(\partial_\mu \phi_{j]}) B^{\mu\nu} = \phi_{[i}\partial_\nu \phi_{j]} \partial_\mu B^{\mu\nu} = \phi_{[i}\partial_\nu \phi_{j]} J_Y^\nu
}
where $J_Y^\nu = \sum_F \mathcal{Y}_F \bar{\Psi}_F \gamma^\nu \Psi_F$ is the hypercharge current.
This mediates a non-zero rate, via an off-shell $B$.
Therefore, to estimate axion production rates, we use the form
\ba{
\sum_{ij}\dfrac{c^{ij}_{aB}}{f_a^2}(\partial_\mu \phi_i)(\partial_\nu \phi_j) B^{\mu\nu} 
= - \sum_{F,i,j}\dfrac{c^{ij}_{aB}}{f_a^2}{\cal Y}_F \phi_{[i}\partial_\nu \phi_{j]} \bar{\Psi}_F \gamma^\nu \Psi_F.
} 
Similar to the Higgs coupling, the production of the $i$-th axion from $\Psi_F, \bar{\Psi}_F$ annihilation is proportional to 
\es{eq:lamab_i}{
\sum_{j\neq i} |c_{aB}^{ij}|^2 \equiv |\lambda_{aB}^i|^2.
}
Accordingly, we remain agnostic about the form of the matrix $c_{aB}^{ij}$ and just specify the vector $\lambda_{aB}^i$.
We note that the symmetric dimension-6 operator is obtained in the IR in the KSVZ and DFSZ axion models~\cite{Biekotter:2023mpd}, but the skew-symmetric operator is zero if $\N=1$.

Given that the dimension-6 operators generally couple all axion states to the SM, we do not consider operators of dimension seven or higher. If dimension-6 operators are suppressed, however, dimension-7 or higher operators could be the leading production channels for some axion states, albeit more suppressed by powers of $f_a$. We summarize the number of independent axion states available at each dimension in Tab.~\ref{tab:N}. \journal{In our results we always assume only minimal couplings to gravity, but allowing for nonminimal couplings leads to additional terms~\cite{Alexander:2025olg}. However, after canonically normalizing the graviton, these terms are additionally Planck suppressed and effectively $d>6$.}

\begin{table}[t]
\centering
\begin{tabular}{cc}
\toprule
$d$ & ${\cal N}_{\rm ind}$ \\
\midrule
5 & $\min({\cal N}, 44)$ \\
6 & ${\cal N}$ \\
\bottomrule
\end{tabular}
\caption{\label{tab:N} The number of independent axion states, ${\cal N}_{\rm ind}$, coupled to the SM at dimension $d$ in the EFT~\eqref{eq:Lagrangian}. Given a specific UV model, ${\cal N}_{\rm ind}$ may be smaller.}
\end{table}

\subsection{Neutrino Masses}
\label{sec:neu_mass}

In our fiducial setup, we set the SM left-handed neutrino masses to zero. When neutrinos are massless, one can remove the diagonal components of $\cl$ because
\es{}{
\dfrac{\partial_\mu a}{f}J_{L,n}^\mu = - \dfrac{a}{f} (\partial_\mu J^\mu_{L,n}) = - \cl^{nn} \dfrac{\alpha_2}{8\pi}\dfrac{a}{f} \WW + \cl^{nn} \dfrac{\alpha_1}{8\pi}\dfrac{a}{f} \BB
}
lets us express that operator as a linear combination of other operators already present in the Lagrangian. 
In this section, we extend the above to non-zero neutrino mass, showing that the additional freeze-in axion abundance is small and justifying our approximation.

\subsubsection{Dirac Masses}

If the SM neutrino masses are Dirac, there are three right-handed neutrino states. To our EFT we can add the operators
\es{}{
\mathcal{L} \supset \dfrac{\partial_\mu a_\nu^{nm}}{f_a}\bar{\nu}_R^n c_{\nu,nm}\gamma^\mu\nu_R^m - (\bar{L} \Tilde{H} Y_\nu \nu_R + {\rm h.c.}).
}
In the above basis, the diagonal entries of ${\bf c}_L$ are in general non-zero.
The first operator adds nine additional axion states coupled to the right-handed neutrinos. The second operator breaks the three $U(1)_{L,n}$ symmetries down to the total lepton number $U(1)_L$. Then the total number of independent dimension-5 operators is $\underbrace{3}_{\text{gauge}}+\underbrace{1}_\text{Higgs}+\underbrace{6 \times 9}_\text{fermion}-\underbrace{3}_\text{redundant} = 55$. 
However, the production rate for the nine additional axions $a_\nu^{nm}$ coupled to right-handed neutrinos is suppressed by the extremely small $Y_\nu \sim 10^{-12}$ (see Sec.~\ref{sec:fermion}). 

The two additional axion states arising from the breaking of $U(1)_{L_e}\times U(1)_{L_\mu} \times U(1)_{L_\tau} \to U(1)_L$ are similarly suppressed by $Y_\nu$ and does not contribute to $\Dneff$. Indeed, if we rotate the diagonal elements of $\cl$ to zero as in Eq.~\eqref{eq:GSrotations}, we get \es{}{
\mathcal{L} \supset - \dfrac{a_L^{nn}}{f_a} (\bar{L}^n \Tilde{H} i c_{L,nn} Y_{\nu,nm} \nu_R^m + {\rm h.c.})
}
where we work in the usual weak-interaction basis where $Y_e$ is diagonal. Using the total $U(1)_L$ symmetry we can remove the trace of $\cl$, so these additional Lagrangian terms correspond to only two additional axions beyond the assumption of massless neutrinos.
Finally, we note the contribution to $\Dneff$ from light right-handed neutrinos themselves is unobservably small~\cite{Luo:2020fdt}, unless there are additional neutrino self-interactions.

\subsubsection{Majorana Masses}

For SM neutrino Majorana masses, we need add no states to the theory, but instead the dimension-5 Weinberg operator
\es{}{
\mathcal{L} \supset (\bar{L}\Tilde{H})\dfrac{{\bf c}_5}{\Lambda}(\bar{L}\Tilde{H}) + {\rm h.c.}.
}
where ${\bf c_5}$ is a complex symmetric $3\times3$ matrix in generation space, and to match the observed masses we have $\Lambda/|{\bf c}_5| \sim 10^{15}$ GeV. Majorana masses break the $U(1)_{L_e}\times U(1)_{L_\mu} \times U(1)_{L_\tau} \to U(1)_L$ with no surviving $U(1)_L$, so there will be 47 independent dimension-5 operators. The additional three axion states interact with the SM suppressed by the scale of L-violation $\Lambda$. Indeed, under the L-rotations of Eq.~\eqref{eq:GSrotations}, we generate a dimension-6 operator
\es{eq:ax_majorana}{
\mathcal{L} \supset \sum_{m,n}(\bar{L}_n\Tilde{H})\dfrac{c_{L,nn} c_{5,nm} a_L^{nn} - c_{5,nm} c_{L,mm} a_L^{mm}}{\fa \Lambda}(\bar{L}_m\Tilde{H}) + {\rm h.c.}.
}
The simplest UV completions ({\it e.g.,} see-saw~\cite{Yanagida:1979as,Gell-Mann:1979vob,Glashow:1979nm,Ramond:1979py}) require heavy right-handed neutrinos above the scale $\Lambda$. A careful evaluation of the associated axion production in that case is beyond the scope of this work; nevertheless, we note that our results only strictly hold for $\trh,\fa<\Lambda$ if the neutrino masses are Majorana-type. Below this scale, the axion production from Eq.~\eqref{eq:ax_majorana} can be ignored since the interaction is suppressed by $\Lambda f_a/|{\bf c}_5|$, as opposed to $f_a^2$, and $\Lambda/|{\bf c}_5| > f_a$ for almost all of our parameter space.
Furthermore, there is a kinematic suppression due to the quintic interaction and the three-body final-state phase space. Therefore we expect that additional axion production is highly suppressed.

\section{Benchmark EFT Couplings}
\label{sec:benchmark}

In this section, we consider several benchmark scenarios for axion couplings to the SM, which strongly differ in their predictions for $\Dneff$. We assume that there are $\N$ axions, where $\N \sim \mathcal{O}(1-100)$ or larger.
These axions will thermalize in the early universe if their interaction rates with the SM are ever above the Hubble rate. As we will show, the dimension-5 interactions could thermalize as many as 44 axions (if $\N\geq 44$), while the dimension-6 operator could thermalize all $\N$ axions.
The resulting $\Dneff$ depends on the couplings of the axions to the SM and the reheating temperature $\trh$. 
These results for $\Dneff$, within each benchmark EFT, are shown in Sec.~\ref{sec:result}.
We first analyze hadronic axions, which couple only to gauge bosons, and then axions which additionally interact with the SM fermions, under various assumptions for the flavor structure of those couplings.

\subsection{Hadronic Axion Couplings}

In one of the most minimal scenarios, the axions do not possess tree-level couplings to the SM fermions. Such a setup is realized in the context of hadronic axion models, {\it e.g.}, the KSVZ axion~\cite{Kim:1979if,Shifman:1979if}. A simple UV completion yielding this scenario is $\N$ independent copies of the KSVZ axion, such that there is a $(U(1)_{\rm PQ})^\N$ PQ symmetry which is spontaneously broken to unity at the scale $f$.\footnote{While this UV completion may seem less appealing when ${\cal N}\gg 1$, in theories with several compact extra dimensions, ${\cal N}\gg 1$ appears naturally thanks to the many ways of wrapping higher form fields around the compact dimensions.} Below the joint PQ-breaking scale, the $\N$ axions acquire couplings to gauge bosons and to the Higgs through the dimension-6 operator, but not to the fermions at tree-level. 
The EFT can then be written as
\es{}{
\mathcal{L} \supset \sum_i c^i_G \dfrac{\alpha_s}{8\pi} \dfrac{\phi_i}{f} \GG + \sum_i c^i_2 \dfrac{\alpha_2}{8\pi} \dfrac{\phi_i}{f} \WW + \sum_i c^i_1 \dfrac{\alpha_1}{8\pi} \dfrac{\phi_i}{f} \BB + \sum_{ij}\dfrac{c_{aH}^{ij}}{f^2}(\partial_\mu \phi_i)(\partial^\mu \phi_j)|H|^2.
}
In fact, this $(U(1)_{\rm PQ})^\N$ UV-completion yields a diagonal $c_{aH}$, from integrating out the associated radial modes, but we work with the more general Lagrangian not assuming any specific UV physics. 
We then follow the procedure discussed in Sec.~\ref{sec:axEFT} to go to a basis such that only one axion appears in each dimension-5 operator:
\es{eq:had_ax}{
\mathcal{L} \supset \dfrac{\alpha_s}{8\pi} \dfrac{\aqcd}{f_a} \GG + c_2 \dfrac{\alpha_2}{8\pi} \dfrac{a_2}{f_a} \WW + c_1 \dfrac{\alpha_1}{8\pi} \dfrac{a_1}{f_a} \BB + \textrm{[dim. 6]}.
}
As mentioned in Sec.~\ref{sec:axEFT}, the Gram-Schmidt procedure will also generate couplings of $\aqcd$ to $\WW$ and $\BB$, and $a_2$ to $\BB$.
These are not included here since we are interested in a conservative, lower bound on the freeze-in yield.
Therefore, at dimension five, only these three physical axion states, $\aqcd$, $a_2$, and $a_1$ interact with the SM, while all other axion states are sterile, so that the largest possible contribution to $\Dneff \approx 3 \times 0.027$, for decoupling temperature larger than the electroweak scale. This remains true even if new axion couplings, for example, to SM fermions, are generated
by operator mixing at one loop,
because those operators will only couple to axions which are linear combinations of $\aqcd$, $a_2$, and $a_1$.

\paragraph{Grand Unified Theory Couplings.}
The above three axions need not be independent. For example, if the SM is embedded in a GUT with gauge group $G_{\rm GUT}$, then we get a single linear combination that has an anomalous coupling to the unified gauge group $G_{\rm GUT}$ at dimension five~\cite{Srednicki:1985xd} (see also~\cite{Agrawal:2022lsp} for a recent work):
\es{eq:gut_ax}{
\mathcal{L} \supset \dfrac{\alpha_{\rm GUT}}{8\pi f_a} a_{\rm GUT} G_{\rm GUT}\tilde{G}_{\rm GUT} + \textrm{[dim. 6]}.
}
The GUT-axion $a_{\rm GUT}$ plays the role of the QCD axion.
If it thermalizes, we expect the standard $\Dneff \approx 0.027$ result for one additional scalar decoupled above the weak scale.
Depending on the value of $f_a$, there can also be a sizable non-relativistic population of $a_{\rm GUT}$, either from misalignment or topological defect production.

\subsection{Couplings to Fermions}

In this section, we consider an EFT where axions couple to fermions at tree-level, in addition to the gauge sector, motivated by the DFSZ scenario~\cite{Dine:1981rt,Zhitnitsky:1980tq}. Those couplings can be written as in~\eqref{eq:Lagrangian}, which we reproduce here for convenience,
\es{eq:ferm_ax}{
{\cal L} \supset \dfrac{\alpha_s}{8\pi} \dfrac{a_{\rm QCD}}{f_a} \GG + c_B \dfrac{\alpha_1}{8\pi} \dfrac{a_B}{f_a} \BB + \sum_{F,n,m \neq L,i,i} \Psibar_F^n c_{F,nm} \dfrac{\partial_\mu a_F^{nm}}{f_a}\gamma^\mu\Psi_F^m,
}
for the SM multiplets $F \in Q,L,u,d,$ and $e$. Here, ${\textbf c}_F$ is a $3\times3$ Hermitian matrix, where the entries span the three generations (indexed by $m$ and $n$) so that its off-diagonal elements correspond to flavor-violating couplings. Note that this class of operator is invariant under CP\@. 
As discussed in Sec.~\ref{sec:axEFT}, some of the fermion couplings can be removed by doing field redefinitions.
We will work in a basis such that diag$({\bf c}_L) = (0,0,0)$ and $c_2=0$ (no coupling to $SU(2)_L$) for simplicity, although our results do not depend on this specific choice. In the following subsections, we calculate the number of axion degrees of freedom that have nonzero couplings to the SM fermions under various assumptions for the flavor structure of these couplings.

\subsubsection{Anarchy}
\label{sec:anarchy}
We first make the assumption that all entries in ${\textbf c}_F$ are anarchic~\cite{Hall:1999sn}, meaning that they are complex random variables. 
We see that at dimension five, there are 44 axion degrees of freedom, taking into account that diag$({\bf c}_L)=0$, that couple to the SM\@. Our prior on the Wilson coefficients is such that each axion couples with $\mathcal{O}(1)$ strength.
In practice, in our results, we set each of the entries of the ${\bf c}_F$ matrix to have unit magnitude and arbitrary uncorrelated phases, i.e., 
\es{}{\label{eq:c_Anarchy}
|c_{F,nm}| = 1~{\rm for}~F=Q,u,d,L,e,
}
subject to diag$({\bf c}_L)=0$ and ${\bf c}_F^\dagger = {\bf c}_F$.
If all axions thermalized with the SM at high temperatures, the resulting $\Dneff\geq 0.027\times44\approx1.17$, a scenario which is excluded at high significance. In the next sections, we compute the decoupling temperature for each axion and show that, while large reheating temperatures are in tension with existing constraints, low reheating scenarios are viable, and there is a large range of $T_{\rm RH}$ which could lead to signals at upcoming CMB observatories.

\subsubsection{Froggatt-Nielsen Texture}
\label{sec:FN}
We next assume that the axion couplings to the fermion sector are such that the SM flavor puzzle is solved by the Froggatt-Nielsen (FN) mechanism~\cite{Froggatt:1978nt} at a scale $\lfn \gg f_a$. In the FN setup, the SM is extended with a global $U(1)_{\cal H}$ horizontal symmetry which is spontaneously broken by the VEV $\langle S\rangle$ of a new complex scalar $S$. The SM Yukawas arise solely from powers of $\epsilon \equiv \langle S\rangle/\lfn$.

This idea can be generalized by considering a {\it texture}~\cite{Leurer:1992wg, Leurer:1993gy}, the set of charges of the SM fermions under $U(1)_{\cal H}$.
In this work, we consider such an example of a texture, but we take a bottom-up approach and do not specify the UV degrees of freedom that provide such charges.
We adopt a convention where $S$ has charge 1 and the SM Higgs has charge 0 under $U(1)_{\cal H}$, we can express the SM Yukawa couplings
\es{eq:sm_yukawa}{
\mathcal{L} \supset Y_{d, nm} \bar{Q}_n H d_m + Y_{u, nm}  \bar{Q}_n \tilde{H} u_m + {\rm h.c.},
}
where $\tilde{H} = i\tau_2 H^\dagger$, as
\es{}{
\mathcal{L} \supset \kappa_{d, nm}\left({S\over \Lambda_{\rm FN}}\right)^{n_{nm}^d} \bar{Q}_n H d_m + \kappa_{u, nm}\left({S\over \Lambda_{\rm FN}}\right)^{n_{nm}^u} \bar{Q}_n \tilde{H} u_m + {\rm h.c.}\\ \to \kappa_{d, nm}\epsilon^{n_{nm}^d} \bar{Q}_n H d_m + \kappa_{u, nm}\epsilon^{n_{nm}^u} \bar{Q}_n \tilde{H} u_m + {\rm h.c.},
}
where $\kappa$'s are some ${\cal O}(1)$ numbers.
In most implementations, the ratio $\langle S\rangle/\lfn$ is identified with $\epsilon\equiv\sin\theta_c\approx0.23$, and a texture is identified that reproduces the SM Yukawas at low energies. To conserve charge, we see that $n_{nm}^d = |{\cal H}(Q_n)-{\cal H}(d_m)|$ and $n_{nm}^u = |{\cal H}(Q_n)-{\cal H}(u_m)|$. 
Note that if the argument of the absolute value is negative, then instead $S^\dagger$ is coupled to the SM fermions. 
Various choices of texture are possible, see, {\it e.g.},~\cite{Leurer:1992wg, Leurer:1993gy}.
Once the FN charges are assigned to the SM fermions, they also dictate the sizes of the couplings of those fermions to other fields, including new physics scenarios like the axion couplings. In particular, the structure of the $\phi_i$ coupling to the SM fermions requires $c_{F,nm} \propto \epsilon^{|{\cal H}(\Psi_F^n)-{\cal H}(\Psi_F^m)|}$.
Using the horizontal charge assignments given in Table~1 of~\cite{Asadi:2023ucx}, we find 
\es{}{
c_Q &=
    \begin{pmatrix}
\epsilon^0 & \epsilon^{|X|} & \epsilon^{|3X|}\\
\epsilon^{|X|} & \epsilon^0 & \epsilon^{|2X|}\\
\epsilon^{|3X|} & \epsilon^{|2X|} & \epsilon^0\\
\end{pmatrix},
c_u =
    \begin{pmatrix}
\epsilon^0 & \epsilon^{|4X\pm 7|} & \epsilon^{|3X\pm 7|}\\
\epsilon^{|4X\pm 7|} & \epsilon^0 & \epsilon^{|X|}\\
\epsilon^{|3X\pm 7|} & \epsilon^{|X|} & \epsilon^0\\
\end{pmatrix},
c_d =
    \begin{pmatrix}
\epsilon^0 & \epsilon^{|6X\pm 6|} & \epsilon^{|5X\pm 6|}\\
\epsilon^{|6X\pm 6|} & \epsilon^0 & \epsilon^{|X|}\\
\epsilon^{|5X\pm 6|} & \epsilon^{|X|} & \epsilon^0\\
\end{pmatrix},\\
c_L &=
    \begin{pmatrix}
\epsilon^0 & \epsilon^{|Y|} & \epsilon^{|Y|}\\
\epsilon^{|Y|} & \epsilon^0 & \epsilon^0\\
\epsilon^{|Y|} & \epsilon^0 & \epsilon^0\\
\end{pmatrix},
c_e =
    \begin{pmatrix}
\epsilon^0 & \epsilon^{|6Y\mp 8|} & \epsilon^{|4Y\mp 8|}\\
\epsilon^{|6Y\mp 8|} & \epsilon^0 & \epsilon^{|2Y|}\\
\epsilon^{|4Y\mp 8|} & \epsilon^{|2Y|} & \epsilon^0\\
\end{pmatrix}.
}
The numbers $X,Y$ take values $\pm 1$ and together with the other sign choices in $c_u, c_d, c_e$, there are $2^5$ possibilities for the FN charges.
For an example benchmark, we choose $X=-Y=-1$, the positive sign in $c_u, c_d$, and the negative sign in $c_e$; this choice is motivated by holomorphy of a SUSY UV completion of the axiverse EFT~\cite{Asadi:2023ucx}.
This leads to
\es{eq:texture}{
c_Q &=
    \begin{pmatrix}
\epsilon^0 & \epsilon^{1} & \epsilon^{3}\\
\epsilon^{1} & \epsilon^0 & \epsilon^{2}\\
\epsilon^{3} & \epsilon^{2} & \epsilon^0\\
\end{pmatrix},
c_u =
    \begin{pmatrix}
\epsilon^0 & \epsilon^{3} & \epsilon^{4}\\
\epsilon^{3} & \epsilon^0 & \epsilon^{1}\\
\epsilon^{4} & \epsilon^{1} & \epsilon^0\\
\end{pmatrix},
c_d =
    \begin{pmatrix}
\epsilon^0 & \epsilon^{0} & \epsilon^{1}\\
\epsilon^{0} & \epsilon^0 & \epsilon^{1}\\
\epsilon^{1} & \epsilon^{1} & \epsilon^0\\
\end{pmatrix},\\
c_L &=
    \begin{pmatrix}
\epsilon^0 & \epsilon^{1} & \epsilon^{1}\\
\epsilon^{1} & \epsilon^0 & \epsilon^0\\
\epsilon^{1} & \epsilon^0 & \epsilon^0\\
\end{pmatrix},
c_e =
    \begin{pmatrix}
\epsilon^0 & \epsilon^{2} & \epsilon^{4}\\
\epsilon^{2} & \epsilon^0 & \epsilon^{2}\\
\epsilon^{4} & \epsilon^{2} & \epsilon^0\\
\end{pmatrix}.
}

Relative to the anarchic case, the flavor-violating axion couplings are now parametrically suppressed by powers of $\epsilon$. However, this does not change the number of independent degrees of freedom coupled to the SM, so if all axions took on a thermal abundance, we would have $\Dneff\geq 0.027\times44$. The Wilson coefficients are as above, up to an $\mathcal{O}(1)$ coefficient in each entry and a random phase in the off-diagonals.

Finally, we note that the angular mode of $S$ is a light pseudoscalar which could also freeze-in. In this work, we do not consider its contribution to $\Dneff$. In some models it is identified with the axion~\cite{Ema:2016ops,Calibbi:2016hwq}.

\subsubsection{Minimal Flavor Violation}\label{sec:MFV}

Our final assumption for the axion-fermion couplings does introduce correlations between the axions coupled to different SM fermions, so we will now find that there are less than 42 axion-fermion couplings at dimension five. We investigate in this context the principle of Minimal Flavor Violation (MFV)~\cite{DAmbrosio:2002vsn}, where the SM Yukawas transform as spurions under the $SU(3)^5$ flavor symmetry group. Then the axion-fermion couplings can be expanded in powers of the spurions, where $c_F$ is required to transform as $\mathbf{8}_F$ under $SU(3)_\psi \subset SU(3)^5$. It is useful to define the Hermitian matrices $X_U \equiv Y_u Y_u^\dagger \sim \mathbf{8}_Q$, $X_D \equiv Y_d Y_d^\dagger \sim \mathbf{8}_Q$, $U_k \equiv Y_u^\dagger X_D^{k-1} Y_u \sim \mathbf{8}_u$, $D_k \equiv Y_d^\dagger X_U^{k-1} Y_d \sim \mathbf{8}_d$, $L \equiv Y_e Y_e^\dagger \sim \mathbf{8}_L$, $E \equiv Y_e^\dagger Y_e \sim \mathbf{8}_e$. Using the Cayley-Hamilton theorem, we expand the couplings in only these matrices and remove related terms. We find
\es{eq:MFV}{
c_Q =&\, c_Q^{(0)} \mathit{I} + c_Q^{(1)}X_U + c_Q^{(2)}X_D + \mathcal{O}(Y^4) \\
c_u =&\, c_u^{(0)} \mathit{I} + c_u^{(1)}U_1 + \mathcal{O}(Y^4) \\
c_d =&\, c_d^{(0)} \mathit{I} + c_d^{(1)}D_1 + \mathcal{O}(Y^4) \\
c_L =&\, c_L^{(0)} \mathit{I} + c_L^{(1)}L + c_L^{(2)}L^2 \\
c_e =&\, c_e^{(0)} \mathit{I} + c_e^{(1)}E + c_e^{(2)}E^2
}
where all expansion coefficients are expected to be $\mathcal{O}(1)$. In the limit of zero neutrino masses, the expansion in the lepton sector is complete. 
While we have not explicitly written the full expansion in the quark sector, it can be systematically computed~\cite{Colangelo:2008qp}.
In our numerical computation, we use the full expansion without the truncation shown above.
It is useful to work in a basis where one of the quark Yukawa couplings is diagonal. We label the up-basis the one where $Y_u = {\rm diag}(y_u,y_c,y_t)$ and $Y_d = \vckm{\rm diag}(y_d,y_s,y_b)$. The down-basis is related by a $SU(3)_Q$ rotation by $\vckmd$. Explicitly, in either basis $c_u$, $c_d$, $c_L$, and $c_e$ are identical, while $c_Q^{\rm up} = \vckm c_Q^{\rm down}\vckmd$. We choose to work in the up-basis. We have explicitly checked that working in the down-basis leads to identical results.

Each term in the expansion~\eqref{eq:MFV} generically leads to independent axion states, because each $c_F^{(i)}$ is a random uncorrelated $\mathcal{O}(1)$ number. However, the spurion expansion parameters are highly aligned in the space of $3\times3$ Hermitian matrices, meaning that the axion states thermalized by bath interactions can be highly aligned. To ensure we are working with orthogonal axion state vectors, we apply the Gram-Schmidt orthogonality procedure to the above expansion, ordering the vectors as in the MFV expansion. That is, for each $F$ we first define an axion state $a_F^{(0)}$ which couples flavor-universally (through the identity matrix $\mathit{I}$) with strength $c_F^{(0)}$. We then define a second axion state $a_F^{(1)}$, which couples to flavor suppressed by an ${\bf 8}_F$ spurion, but with $a_F^{(0)}$ projected out. We iteratively continue the procedure for all the terms in~\eqref{eq:MFV}. We find that the only axion states which have nonnegligible freeze-in abundance are those which are flavor-universal, or $a_Q^{(1)}$ and $a_u^{(1)}$, which couple mostly to the top quark. We note one could have obtained this same result by setting all but the top Yukawa to zero and the CKM matrix diagonal from the beginning. Explicitly, to an extremely good approximation, we then have

\es{eq:mfv}{
c_Q &= c_Q^{(0)}\mathit{I} + c_Q^{(1)} y_t^2/3\, {\rm diag}(-1,-1,2) \\
c_u &= c_u^{(0)}\mathit{I} + c_u^{(1)} y_t^2/3\, {\rm diag}(-1,-1,2)\\
c_d &= c_d^{(0)}\mathit{I}\\
c_L &= c_L^{(0)}\mathbf{0}\\
c_e &= c_e^{(0)}\mathit{I},
}
recalling that the diagonal elements of $c_L$ are set to zero in the basis we are working in.

\section{Axiverse Cosmology}\label{sec:computation}
So far we have discussed the EFT couplings of $\N$ axions with the SM\@.
One common presence in these EFTs is the QCD axion $\aqcd$, which also solves the strong CP problem in these scenarios.
In this section, we discuss the cosmology of $\aqcd$, as well as that of all the remaining axions, and justify when we can treat these remaining axions effectively as massless.
Throughout, we assume a standard radiation-dominated cosmology below the reheat temperature $\trh$. Our results for $\Dneff$ can be modified straightforwardly if there is additional entropy injection, {\it e.g.}, due to transient epochs of early matter domination.  Our bounds can also be applied conservatively to scenarios with early matter domination, identifying $\trh$ with the temperature of the SM plasma at the start of the final radiation dominated era.
\subsection{Effects of Mass Mixing}\label{sec:ax_mass}
In general, the mass mixing between different axions can have an impact on their cosmological abundance.
This is because the `mass basis' and the `flavor basis' need not be identical.
Therefore, `active' axions produced via a set of SM interactions can `mix' into otherwise `sterile' states.
If that happens efficiently enough at high temperatures, the total $\Delta N_{\rm eff}$ could be larger than expected based on the operator counting given in the previous section.
We can obtain a rough estimate of mass mixing as follows.

For simplicity, consider a two-axion system, where the axions have masses $m_1$ and $m_2$.
The mass basis does not coincide with the flavor basis generically, and therefore, the axion of one flavor can oscillate into another flavor.
In analogy with neutrino oscillations in vacuum, the probability of a flavor oscillation in time $t$ is given by $\sin^2(\vartheta)\times  \sin^2((m_2^2-m_1^2)t/(4k))$~\cite{Raffelt:1996wa}, where $\vartheta$ is the rotation angle between the mass basis and the flavor basis, and $k$ is the 3-momenta of both the axions. 
Here, we have treated both axions to be relativistic.
Since we are interested in the production of these axions from the thermal bath, $k \sim T$, the temperature of the SM bath.
Taking $t\sim 1/H$, at $T\gg{\rm GeV}$ and $m_2, m_1\ll\rm{eV}$ (as we will justify shortly),
\es{}{
{m_2^2-m_1^2 \over T H} \ll 1.
}
We then estimate the rate of mixing to be $\Gamma_{\rm mix} \sim (m_2^2-m_1^2)/T$.
The other relevant quantity is the typical scattering rate of axion with the SM bath, which goes as $\Gamma_{\rm scatter} \sim \kappa T^3/f_a^2$, where $\kappa$ is a factor that depends on the precise axion coupling with the SM\@.
Hence,
\es{}{
{\Gamma_{\rm mix}\over \Gamma_{\rm scatter}} \sim {(m_2^2-m_1^2) f_a^2 \over \kappa T^4} \lesssim {m_{\rm QCD}^2 f_a^2 \over \kappa T^4}.
}
In the last step, we have assumed that the heavier axion is the QCD axion with a mass $m_{\rm QCD} \sim m_\pi f_\pi/f_a$.
As long as $\Gamma_{\rm mix}/\Gamma_{\rm scatter}\rvert_{T_d} <1$ at the decoupling temperature $T_d$, the effect of mixing on $\Dneff$ is negligible.
This is because even if mixing effects are relevant below $T_d$, it can only redistribute energy density among the different axion species, but cannot `extract' energy from the SM bath.
Therefore, $\Delta N_{\rm eff}$ does not change due to mixing effects, assuming all the axions remain relativistic.
The decoupling temperature can be roughly estimated as $T_d \sim f_a^2/(\kappa\mpl)$ (see below for a more precise expression).
Therefore, the previous condition implies that mixing effects can be neglected when
\es{}{
f_a > (m_\pi f_\pi)^{1/4} \mpl^{1/2}\kappa^{3/8} \sim 7.4\times 10^6~{\rm GeV}\left({\dfrac{\kappa}{10^{-5}}}\right)^{3/8},
}
where we have normalized using $\kappa \sim 10^{-5}$, relevant for the QCD axion.
Because of the running QCD coupling, $\kappa$ is necessarily energy dependent, but taking $\kappa\sim 10^{-5}$ provides a reasonable approximation for the above relation in the parameter space of our interest.
This lower bound on $f_a$ is weaker than the astrophysical and rare meson decay bounds discussed later.
This implies that for the observationally viable parameter space, we can ignore the effects of mass mixing.
Alternatively, this condition on $f_a$ implies $m_{\rm QCD}\lesssim 1$~eV, and therefore, as long as all the other axions have a mass smaller than that, we can neglect mass mixing effects in determining $\Dneff$.
This restriction is also consistent with our assumption that all the axions are relativistic during CMB decoupling which also implies all the axions should be lighter than $\sim 0.1$~eV\@. We note that thermal axions oscillating into heavy axions may be a possible DM production mechanism, which we leave to future work.

\subsection{QCD Axion Cosmology}
Since the presence of the QCD axion $\aqcd$ is generic in our EFT description, we first review the associated cosmology.
Throughout this and the following discussion, we will assume a `preinflationary' scenario where the PQ symmetry remains broken during and after inflation.
We will also consider production from the thermal bath, and any additional production channel (for example, in the `postinflationary' scenario) would give a larger $\Dneff$, and in this sense, our derived sensitivity on $\Dneff$ will be conservative.

Depending on its mass, the QCD axion can either contribute as dark matter or dark radiation.
Since our EFT treatment is valid for $T_{\rm RH} < f_a$, we only focus on DM production through the misalignment mechanism, which implies~\cite{ParticleDataGroup:2024cfk}
\es{eq:misal}{
\Omega_{\aqcd}h^2 \approx 0.12 \left({f_a \over 9\times 10^{11}~{\rm GeV}}\right)^{1.16} \theta_i^2,
}
for misalignment angle $\theta_i\lesssim 1$.
The QCD axion can also be produced from the SM bath, either through pion scattering or through quarks and gluons, depending on the temperature.
Ref.~\cite{Bianchini:2023ubu} places an upper bound $m_{\rm QCD}\leq 0.16$~eV at 95\% CL, implying $f_a \geq 3.6\times 10^7$~GeV, by computing $\aqcd$ production from pion scattering at temperatures $\lesssim 150$~MeV, the QCD crossover temperature \journal{(see also Ref.~\cite{DEramo:2022nvb} for an analysis with higher initial temperatures, with similar results).}
\journal{Above this temperature, but below the electroweak scale, perturbative computations are not fully trustworthy. See Refs.~\cite{GrillidiCortona:2015jxo,DEramo:2021psx,DEramo:2021lgb,Notari:2022ffe} for a discussion of the regime of validity of perturbation theory of QCD at finite temperature. However, convergence of perturbative QCD improves at higher temperatures, and we expect the associated rates to be under better control for $T_{\rm RH}\gtrsim$~TeV, which is where we focus on in this work.}

\subsection{Cosmological Abundance of the Remaining Axions}
Similar to the QCD axion, the misalignment mechanism can also produce a significant population of non-QCD axions.
Assuming a standard cosmological evolution, their abundance is given by
\es{}{
\Omega_a h^2 \approx 0.12 \left({\theta_i f_a \over 1.3 \times 10^{12}~{\rm GeV}}\right)^2 \left({m_a \over 0.01~{\rm eV}}\right)^{1/2} \left({3.36 \over g_*(T_{\rm osc})}\right)^{1/4},
}
where $T_{\rm osc}$ is the temperature of the SM bath when the axion starts to oscillate.
Thus, this abundance is sensitively dependent on the mass $m_a$ of the light axions, along with $f_a$ and $\theta_i$.
Knowing the mass spectrum of the light axions requires knowledge of the UV theory.
In the following, we remain agnostic about this and focus on the thermal production of the light axions from the SM bath.

This thermal production is dictated by the appropriate decoupling temperature $T_d$ and the reheat temperature $\trh$.
In particular, if $\trh \gg T_d$, the axion would come into thermal equilibrium with the SM, and could significantly contribute to $\Dneff$ depending on $T_d$.
To estimate $T_d$, we use
\es{}{
\Gamma_a(T=T_d) \simeq H(T=T_d).
}
Here $\Gamma_a$ is the total interaction rate of the axion $a$ with SM species; in particular, an axion may interact through several channels of the form $a \psi_1 \leftrightarrow \psi_2 \psi_3$ with SM fields $\psi_{1,2,3}$, which by assumption are in equilibrium. In that case, the total interaction rate is 
\es{}{
\Gamma_a = \sum_i n_{1,i}^{\rm eq} \langle\sigma v\rangle_i
}
where $i$ indexes the channels, $n_{1,i}^{\rm eq}$ is the equilibrium number density of the field $\psi_1$ and $\langle\sigma v\rangle_i$ is the thermally-averaged cross section for process $i$. The cross section can be written as $\langle\sigma v\rangle_i = \gamma_i/f_a^2$ for some dimensionless constant $\gamma_i$ since all interactions are at dimension five.\footnote{We discuss the case of dimension-6 operators separately.} We then have
\es{}{
\Gamma_a = \dfrac{\zeta(3)}{\pi^2}\dfrac{T^3}{f_a^2} \sum_i \gamma_i \times 
\begin{cases}
    g_i, & \psi_1{\rm\ is\ a\ boson} \\
    3g_i/4, & \psi_1{\rm\ is\ a\ fermion}
\end{cases}
}
where $g_i$ represents the number of degrees of freedom associated with $\psi_1$ in the $i$th process. Defining the sum over $i$ equal to $\gamma$, we find
\ba{
T_d = \dfrac{\pi^3}{\sqrt{90}\zeta(3)\gamma}\sqrt{g_*}\dfrac{f_a^2}{M_{\rm Pl}}.
}
Here $M_{\rm Pl}$ is the reduced Planck mass $1/\sqrt{8\pi G}$, while $g_*$ should be evaluated at $T_d$.
Now if $\trh \gg T_d$, the axion contributes to dark radiation:
\es{eq:dneff_th}{
\Dneff \equiv {8\over 7}\left({11\over 4}\right)^{4/3} {\rho_a \over \rho_\gamma}\bigg\rvert_{\rm CMB} \approx \dfrac{4}{7} \left({11\over 4}\right)^{4/3} \left(\dfrac{g_{*s}^{th}(T_{\rm CMB})}{g_{*s}^{th}(T_d)}\right)^{4/3} \simeq 0.027,
}
where in the last equality we have assumed that $T_d \gg m_t$ so that the axion abundance does not get diluted by SM entropy injection, with $g_{*s}^{th}(T_d) = 106.75$ and $g_{*s}^{th}(T_{\rm CMB})=3.92$, where the $th$ superscript refers to the value of $g_{*s}$ only accounting for species in thermal contact with the SM bath. 
As axions decouple, $g_{*s}$, which enters $T_d$, changes. For allowed parameter space, axions can contribute at most $\sim10$ to $g_{*s}$, so that assuming the SM value $g_{*s}(>{\rm TeV})=106.75$ introduces a $\sim5\%$ error in the calculation of $T_d$. Thus we adopt this assumption in our $T_d$ calculations for axions coupled at dimension five. However, for large ${\cal N}$, $T_d$ can be substantially modified for axions coupled at dimension six and we use $g_{*s} = 106.75 + {\cal N}$.

On the other hand, for $\trh \ll T_d$, the axions do not thermalize with the SM bath.
Therefore, their abundance is appropriately scaled down by the ratio of the interaction rate and the Hubble rate at $\trh$.

\subsection{Summary of Various Rates}\label{sec:rates}
Several processes contribute to axion production from the thermal bath.
It is convenient to organize the following discussion based on the types of axion couplings with the SM\@.
\subsubsection{Production via Gauge Couplings}
\begin{figure}[!htb]
    \centering
    \includegraphics[width=\textwidth]{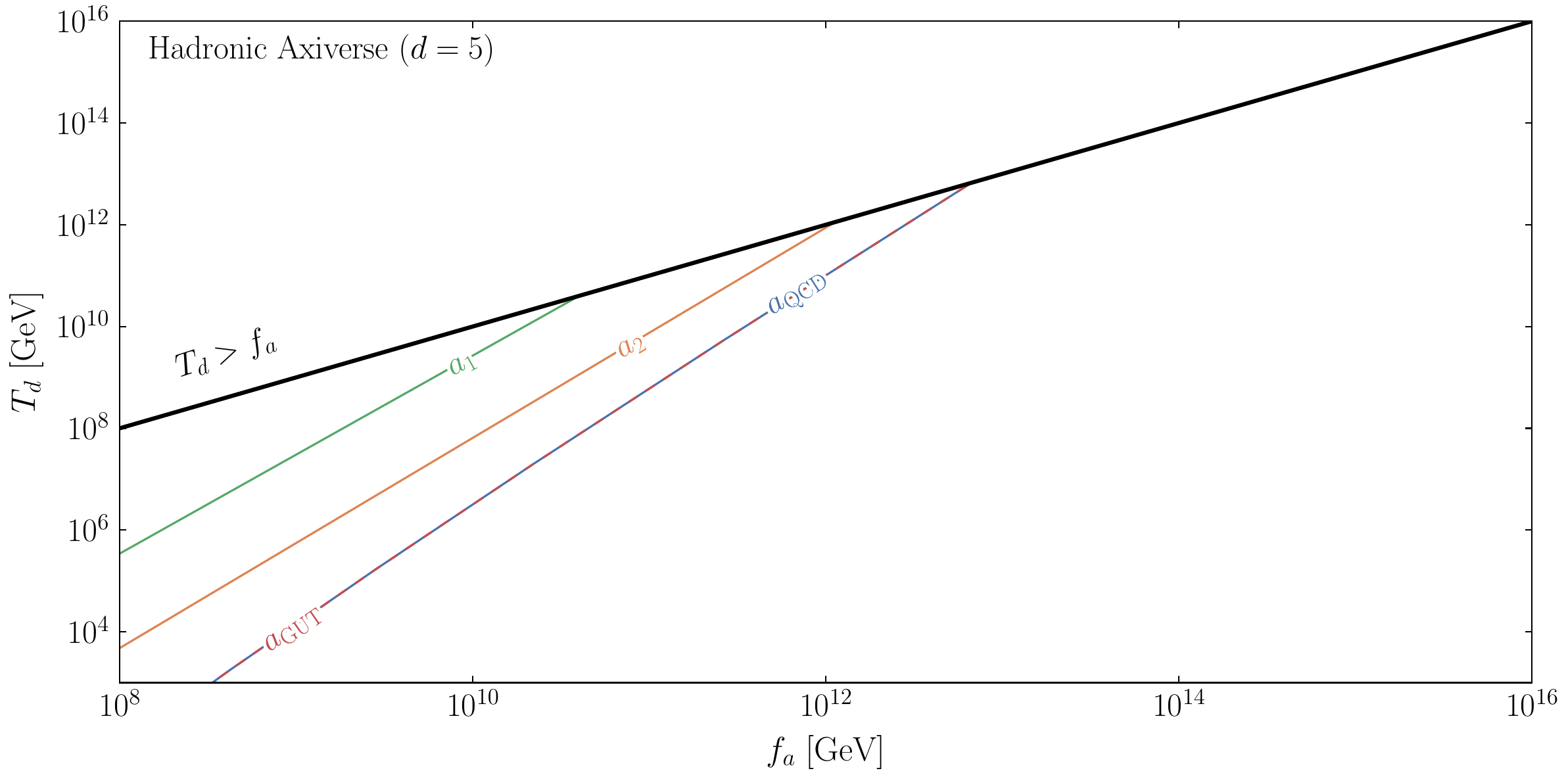}
    \caption{The decoupling temperatures $T_d$ for the independent axion degrees of freedom, assuming that the axion couplings to the SM are hadronic (Eq.~\eqref{eq:had_ax}). There are three independent axions, the QCD axion $a_{\rm QCD}$ (blue), the electroweak axion $a_2$ (orange), and the hypercharge axion $a_1$ (green). We also show the $T_d$ for a scenario where the three axion states arise from a GUT (Eq.~\eqref{eq:gut_ax}), so that there is one independent axion (red). These lines are cut off at $T_d=f_a$ above which our EFT treatment requires inclusion of additional states. 
    }
\label{fig:Hadronic_Tdec}
\end{figure}

To estimate axion production from gauge couplings, we use the production rates based on the thermal field theory calculations in~\cite{Salvio:2013iaa} (for calculations at zero temperature, see~\cite{Masso:2002np}). 
These rates also determine the axion decoupling temperatures $T_d$, which is when the axion production rate equals the Hubble expansion rate.
We evaluate $T_d$ for the three gauge interactions and show the result in Fig.~\ref{fig:Hadronic_Tdec}, including the GUT scenario.
We also use these thermal rates to obtain an approximate abundance of $\Dneff$,
\es{}{
\Dneff \approx 0.027 \times \min\left(1,{\Gamma \over H}\bigg\rvert_{T=T_{\rm RH}}\right).
}
A more accurate evaluation requires solving for the phase space distribution function of the axion within a thermal field theory setup.
We leave such a calculation for future work.

\subsubsection{Production via Fermionic Couplings}
\label{sec:fermion}
\begin{figure}
    \centering
    \includegraphics[width=0.8\textwidth]{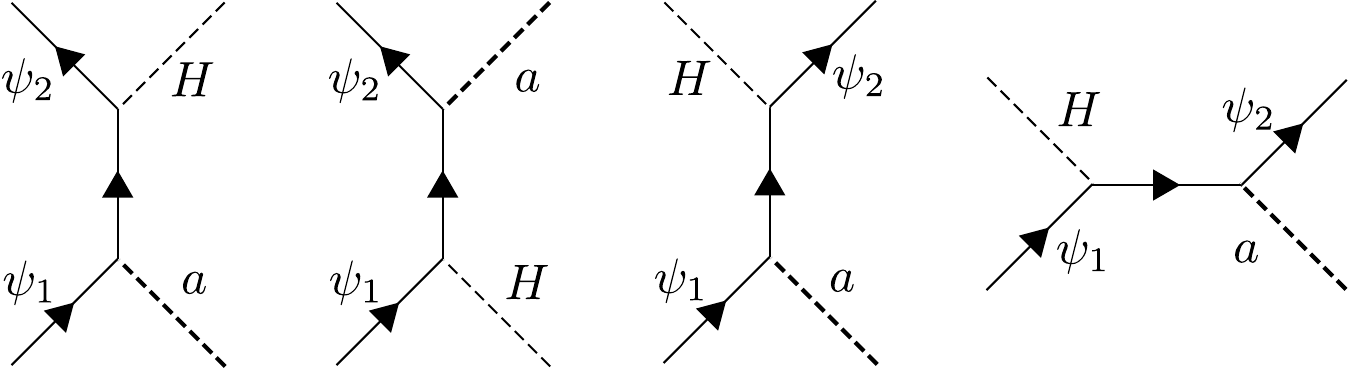}
    \caption{Representative diagrams relevant for freeze-in involving axion-fermion coupling. Since these diagrams are computed at temperatures greater than the weak scale, the full Higgs doublet $H$ participates in the processes. The leading contribution comes from diagrams where the vertex involving $H$ contains the top quark.}
    \label{fig:diag}
\end{figure}

Several diagrams contribute to a freeze-in abundance via axion-fermion couplings, as shown in Fig.~\ref{fig:diag}.
These are the diagrams that do not have a chiral suppression due to a (zero temperature) fermion mass.
The associated freeze-in production can be computed {\it analytically} via solving the full Boltzmann equation for the axion phase space distribution function.\footnote{For a general numerical treatment see, {\it e.g.},~\cite{DEramo:2023nzt, Badziak:2024qjg, DEramo:2024jhn}.}
We can express the relevant processes for freeze-in production as,
\es{}{
\psi_1(p_1)+\psi_2(p_2)\rightarrow \psi_3(p_3) + a(p),
}
where $\psi_i$ are various SM particles, not necessarily all fermions.
The Boltzmann equation for the axion distribution function $\f_a(p,t)$ is given by,
\es{}{
\left({\partial \over \partial t}-H\vecp\cdot \nabla_{\vecp}\right)\f_a(p,t) = {1\over E} C[\f_a],
}
where the collision term, neglecting small quantum statistical effects, is given by
\es{}{
C[\f_a] = {1\over 2}\int \D \Pi_1 \D \Pi_2 \D \Pi_3 (2\pi)^4 \delta^4(p+p_3-p_1-p_2)|{\cal M}|^2(\f_1(p_1,t) \f_2(p_2,t) - \f_3(p_3,t) \f_a(p,t)).
}
Here $\D\Pi_i=\D^3p_i/((2\pi)^3 2E(p_i))$ is the Lorentz invariant phase space integration factor.
The SM particles are in thermal equilibrium at high temperatures: $\f_i = \f_i^{\rm eq}$, and in equilibrium detailed balance gives, $\f^{\rm eq}_1(p_1,t) \f^{\rm eq}_2(p_2,t) = \f^{\rm eq}_3(p_3,t) \f^{\rm eq}_a(p,t)$.
This implies the collision term is 
\es{}{
C[\f_a] &= {1\over 2}\int \D \Pi_1 \D \Pi_2 \D \Pi_3 (2\pi)^4 \delta^4(p+p_3-p_1-p_2)|{\cal M}|^2(\f^{\rm eq}_3(p_3,t) \f^{\rm eq}_a(p,t)) - \f^{\rm eq}_3(p_3,t) \f_a(p,t)),\\
&= {1\over 2}(\f^{\rm eq}_a(p,t)-\f_a(p,t))\int \D \Pi_1 \D \Pi_2 \D \Pi_3 (2\pi)^4 \delta^4(p+p_3-p_1-p_2)|{\cal M}|^2 \f^{\rm eq}_3(p_3,t).
}
Defining a production rate as,
\es{}{
\Gamma(p,t) = \int \D \Pi_1 \D \Pi_2 \D \Pi_3 (2\pi)^4 \delta^4(p+p_3-p_1-p_2)\f_3^{\rm eq} {|{\cal M}|^2\over 2 E},
}
with $E=|\vecp|$,
we can express the full Boltzmann equation as,
\es{}{
{\D \f_a(p,t) \over \D t}=\left({\partial \over \partial t}-H\vecp\cdot \nabla_{\vecp}\right)\f_a(p,t) = (\f_a^{\rm eq}-\f_a) \Gamma(p,t),
}
where we have defined a momentum-dependent `production rate' $\Gamma(p,t)$.
To solve this we note the physical momentum $p=q/a$ where $q$ is the comoving momentum.
We then express $\f_a(p,t) = \tilde{\f}_a(q,a)$ and analogously for the other quantities.
The Boltzmann equation then becomes a collection of ordinary differential equations for each comoving momentum~\cite{DAgnolo:2017dbv},
\es{}{
{\partial \tilde{\f}_a(q,a) \over \partial a} = (\tilde{\f}_a^{\rm eq}(q,a) - \tilde{\f}_a(q,a)){\tilde{\Gamma}(q,a)\over a H}.
}
We consider an equilibrium distribution $\tilde{\f}_a^{\rm eq} = 1/(\exp(q/(aT))-1)$.
Given that we are interested in axion production at temperatures much larger than the weak scale, $g_{*,s}$ remains constant with temperature, implying $T\propto 1/a$ and $\tilde{\f}_a^{\rm eq}$ is time-independent.
Using this, the Boltzmann equation can be solved as
\es{}{
{\tilde{\f}_a(q,a) - \tilde{\f}^{\rm eq}_a(q,a) \over \tilde{\f}_a(q,a_{\rm RH}) - \tilde{\f}^{\rm eq}_a(q,a)} = \exp\left(-\int_{a_{\rm RH}}^a {\D a \over a}{\tilde{\Gamma}\over H}\right).
}
Taking $\tilde{\f}_a(q,a_{\rm RH})\approx 0$, indicating negligible initial abundance,
we can simplify this further
\es{eq:psd_1}{
    \tilde{\f}_a(q,a) \approx \tilde{\f}^{\rm eq}_a(q,a) - \tilde{\f}^{\rm eq}_a(q,a)\exp\left(-\int_{a_{\rm RH}}^a {\D a \over a}{\tilde{\Gamma}\over H}\right).
}
The above expression is general for freeze-in production in 2-to-2 scattering with only one axion in the final state, as long as quantum statistical effects and initial axion abundance are negligible.
In particular, we have not used any specific form of the interaction.

In general, an analytic evaluation of the above integral is not possible.
However, significant simplifications occur if $|{\cal M}|^2$ is only a function of either $\mathsf{s}$ or $\mathsf{t}$, or a simple function of both $\mathsf{s}$ and $\mathsf{t}$, where $\mathsf{s}$ and $\mathsf{t}$ are the standard Mandelstam variables.
This is the case for the processes of interest involving axion-fermion couplings at dimension five at temperatures much larger than the (zero-temperature) SM fermion masses.
In that case the matrix elements of interest are always of the form $|{\cal M}|_i^2 = a_i \mathsf{s}/f^2$ or $|{\cal M}|_i^2 = - a_i \mathsf{t}/f^2$, where $i$ indexes the possible production channels.
Then the integral over the phase space of initial particles becomes trivial.
We use
\es{}{
\int \D \Pi_1 \D \Pi_2 (2\pi)^4\delta^4(p+p_3-p_1-p_2) = \int {\D \Omega_{\rm cm} \over 4\pi} {1 \over 8\pi}\left({2|\vecp_1|\over E_{\rm cm}}\right).
}
Since we are interested in relativistic particles, the last factor in the parenthesis is $\approx 1$. When $|{\cal M}|^2$ is only dependent on $\mathsf{s}$, the angular integration is trivial, while when $|{\cal M}|^2 \propto \mathsf{t}$ the angular integration can still be simply done.
This then leads to,
\es{}{
\Gamma(p,t) = \int \D\Pi_3 \f_3^{\rm eq}{a_i \mathsf{s} \over 16\pi E}
\begin{cases}
    1, & |{\cal M}|_i^2 \propto \mathsf{s} \\
    1/2. & |{\cal M}|_i^2 \propto \mathsf{t}
\end{cases}
}
In the cosmic rest frame and for massless particles, $\mathsf{s} = 2E p_3 (1-\cos\theta)$, where $\theta$ is the angle between $\vecp$ and $\vecp_3$.
For all cases of interest, when $\psi_3$ is a boson, $|{\cal M}|_i^2 \propto s$, and when it is a fermion, $|{\cal M}|_i^2 \propto t$. Therefore, we obtain the production rate to be independent of $p$,
\es{eq:Gamma_dim5}{
\Gamma(p,t) =\tilde{\Gamma}(q,a) = {\zeta(3)\over 16\pi^3} \dfrac{T^3}{f^2} \sum_{\rm channels}\left[\dfrac{3}{8}\right]a_i \equiv {A \zeta(3)\over 16\pi^3} \dfrac{T^3}{f^2},
}
where the quantity in brackets is only present if $\psi_3$ is a fermion and we have defined the $\mathcal{O}(1)$ sum over channels $\equiv A$.
We will compute $A$ in specific scenarios shortly.
Writing $H = \kappa T^2/\mpl$ and using~\eqref{eq:psd_1}, we can obtain the final expression for the axion phase space distribution,
\es{}{
\tilde{\f}_a(q,a) \approx \tilde{\f}^{\rm eq}_a(q,a) - \tilde{\f}^{\rm eq}_a(q,a)\exp\left(-{A \mpl \zeta(3)\over 16\pi^3\kappa}(T_{\rm RH}-T)\right).
}
The energy density in axions is then given by,
\es{}{
\rho_a = \int {\D^3 p \over (2\pi)^3}p \tilde{\f}_a(q,a) = \rho_a^{\rm eq} \left(1-\exp\left(-{A\mpl \zeta(3)\over 16\pi^3\kappa}(T_{\rm RH}-T)\right)\right),
}
with $\rho_a^{\rm eq}=\pi^2 T^4/30$ denoting the energy density in an axion in thermal equilibrium with the bath at temperature $T$.
Defining the decoupling temperature $T_d$ by setting $\Gamma(p,t) = H$ (using~\eqref{eq:Gamma_dim5}) at $T=T_d$ gives, $T_d = 16\pi^3  \kappa/(A \mpl\zeta(3))$.
This gives
\es{eq:rho_dim5}{
\rho_a = {\pi^2 T^4 \over 30}\left(1-\exp\left(-{1 \over T_d}(T_{\rm RH}-T)\right)\right). 
}
Since the above was derived under the assumption that the initial axion number density is negligible, we must have $T\leq \trh \ll T_d$.
In that limit, this gives
\es{}{
\rho_a \approx {\pi^2 T^4 \over 30}{(\trh-T) \over T_d}.
}
Thus, for a given axion-SM coupling at dimension five, knowing $T_d$ lets us compute the axion energy density simply using~\eqref{eq:rho_dim5}.  
We now turn to the question of determining $T_d$ for the three fermionic EFTs discussed in the previous section.

It is simplest to start from a toy model with a left-handed doublet $\mathsf{L}$, a right-handed singlet $\mathsf{r}$, each with three generations labeled by $n$ and $m$, along with distinct axion states $a^{nm}_{\mathsf{L},\mathsf{r}}$,
\ba{
\mathcal{L} \supset \bar{\mathsf{L}}_n c_{\mathsf{L},nm} \dfrac{\partial_\mu a_\mathsf{L}^{nm}}{f_a}\gamma^\mu \mathsf{L}_m + \bar{\mathsf{r}}_n c_{\mathsf{r},nm} \dfrac{\partial_\mu a_\mathsf{r}^{nm}}{f_a}\gamma^\mu \mathsf{r}_m + (\bar{\mathsf{L}}_n H Y^\mathsf{r}_{nm} \mathsf{r}_m + {\rm h.c.}).
}
Here we have written the Lagrangian in the derivative basis, but we have also checked these results in the pseudoscalar basis (see also App.~\ref{app:deriv-pseudo}). 
We compute in the basis where the Yukawa is diagonal. The $\mathsf{L}$-coupled axion $a_{\mathsf{L}}^{nm}$ can be produced through production channels involving the Higgs --- any channel with gauge bosons final states is suppressed by SM fermion masses. We find that the production rate
\es{}{
\Tilde{\Gamma}_{\bar{\mathsf{L}}_n \mathsf{r}_m \to \Hd a_{\mathsf{L}}^{nm}} &= |c_{nm}|^2(Y^{\mathsf{r}}_{mm})^2\dfrac{\zeta(3)}{32\pi^3} \dfrac{T^3}{f_a^2}, \\
\Tilde{\Gamma}_{\mathsf{L}_m\bar{\mathsf{r}}_n \to H a_{\mathsf{L}}^{nm}} &= |c_{nm}|^2(Y^{\mathsf{r}}_{nn})^2\dfrac{\zeta(3)}{32\pi^3} \dfrac{T^3}{f_a^2}, \\
\Tilde{\Gamma}_{H \mathsf{r}_m \to \mathsf{L}_n a_{\mathsf{L}}^{nm}} &= \Tilde{\Gamma}_{H \bar{\mathsf{L}}_n \to \bar{\mathsf{r}}_m a_{\mathsf{L}}^{nm}} = |c_{nm}|^2 (Y^{\mathsf{r}}_{mm})^2 \dfrac{3\zeta(3)}{256\pi^3} \dfrac{T^3}{f_a^2}, \\
\Tilde{\Gamma}_{\Hd \bar{\mathsf{r}}_n \to \bar{\mathsf{L}}_m a_{\mathsf{L}}^{nm}} &= \Tilde{\Gamma}_{\Hd \mathsf{L}_m \to \mathsf{r}_n a_\mathsf{L}^{nm}} = |c_{nm}|^2(Y^\mathsf{r}_{nn})^2\dfrac{3\zeta(3)}{256\pi^3} \dfrac{T^3}{f_a^2}, \\
}
and therefore
\ba{
\Tilde{\Gamma}_{a_\mathsf{L}^{nm}} = |c_{nm}|^2\left((Y^\mathsf{r}_{nn})^2 + (Y^\mathsf{r}_{mm})^2\right)\dfrac{7\zeta(3)}{128\pi^3} \dfrac{T^3}{f_a^2}.
}
Similar arguments yield a production rate for the $\mathsf{r}$-coupled axion $a_\mathsf{r}^{nm}$,
\ba{
\Tilde{\Gamma}_{a_\mathsf{r}^{nm}} = |c_{nm}|^2\left((Y^\mathsf{r}_{nn})^2 + (Y^\mathsf{r}_{mm})^2\right)\dfrac{7\zeta(3)}{128\pi^3} \dfrac{T^3}{f_a^2}.
}

\begin{figure}[!htb]
    \centering
    \includegraphics[width=\textwidth]{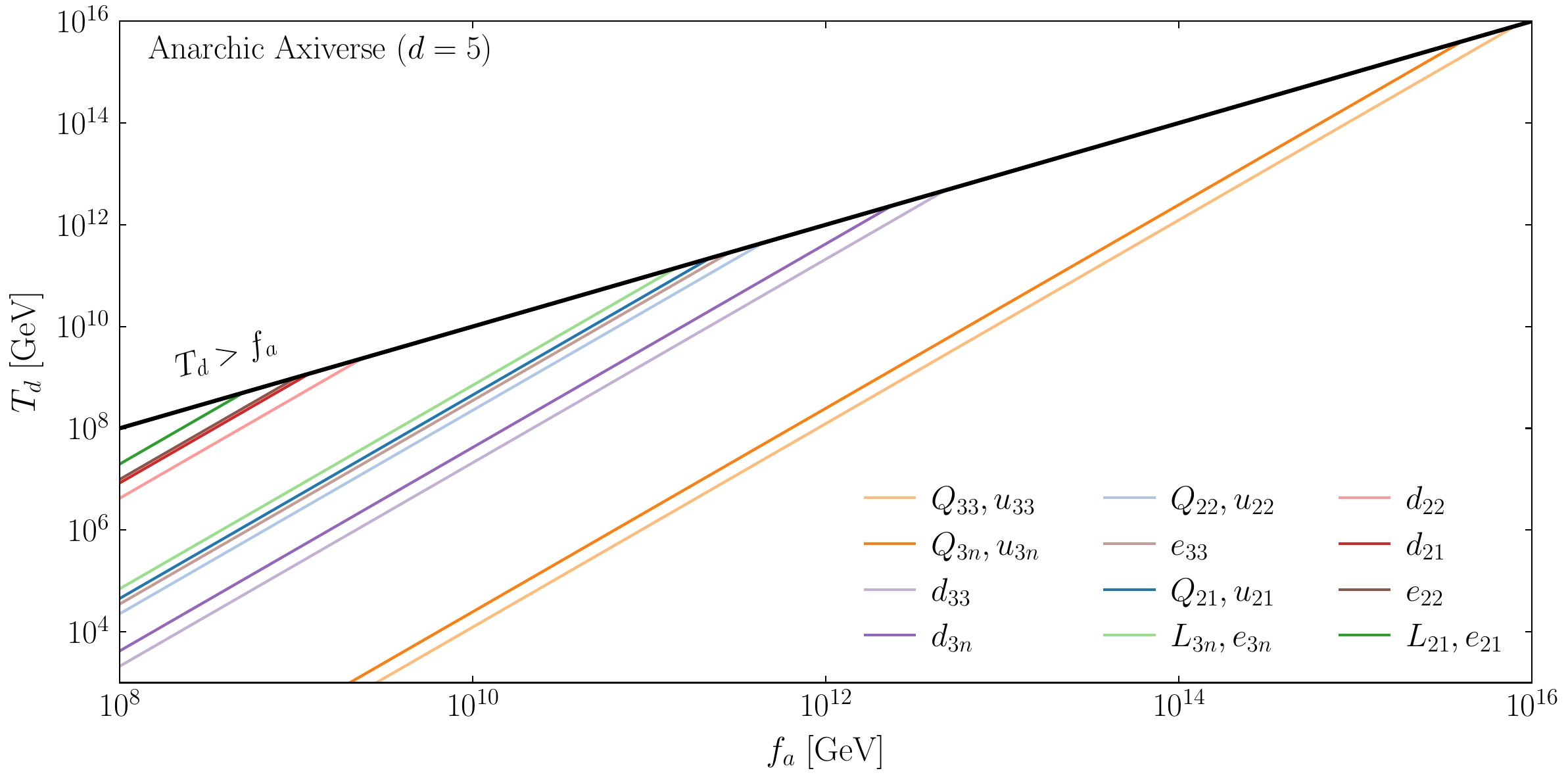}
    \caption{As in Fig.~\ref{fig:Hadronic_Tdec}, but assuming that the axion couplings to fermions are anarchic (Sec.~\ref{sec:anarchy}) with unit magnitudes~\eqref{eq:c_Anarchy}. The legend labels $F_{nm}$ indicate that the associated axion is $a_F^{nm}$ and is listed in ascending order of $T_d$. The subscript $n \in \{1,2\}$ is such that those axions are flavor-violating. The $T_d$ for axions coupled to $F_{ij}$ and $F_{ji}$ is equal, so we only show curves for those with $i>j$. Axion states with no corresponding label have $T_d>f_a$ for all $f_a$ considered (this applies for any state coupled to only first-generation fermions).
    }
\label{fig:Anarchy_Tdec}
\end{figure}

\begin{figure}[!htb]
    \centering
    \includegraphics[width=\textwidth]{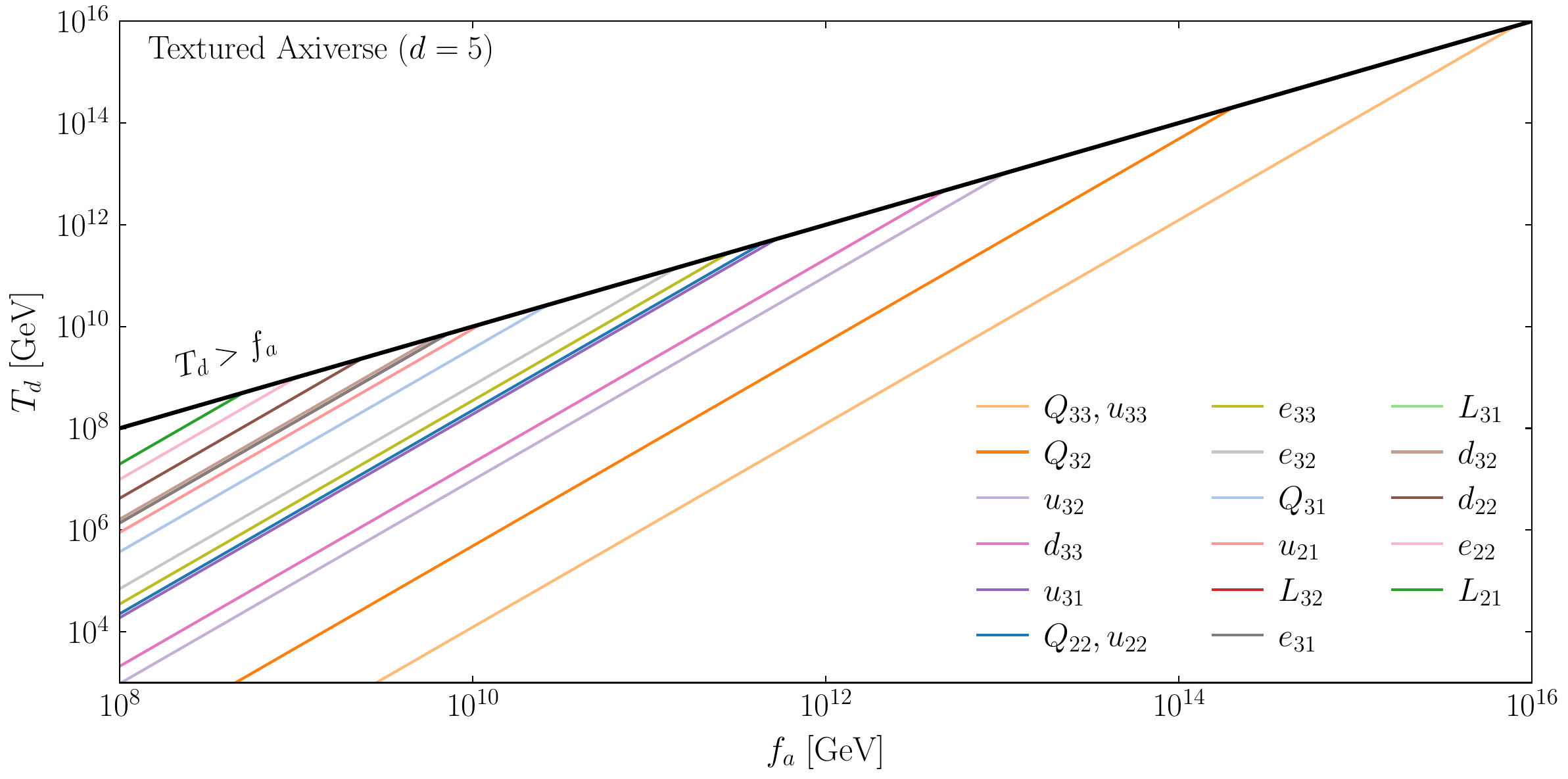}
    \caption{As in Fig.~\ref{fig:Hadronic_Tdec}, but assuming that the axion couplings to fermions are set by a texture (Sec.~\ref{sec:FN}, Eq.~\eqref{eq:texture}). The convention for legend labels is as in Fig.~\ref{fig:Anarchy_Tdec}. Axions are not shown when $T_d>f_a$, including those coupled only to first-generation fermions and to $Q_{21}$, $d_{21}$, and $e_{21}$.
    }
\label{fig:Texture_Tdec}
\end{figure}

\begin{figure}[!htb]
    \centering
    \includegraphics[width=\textwidth]{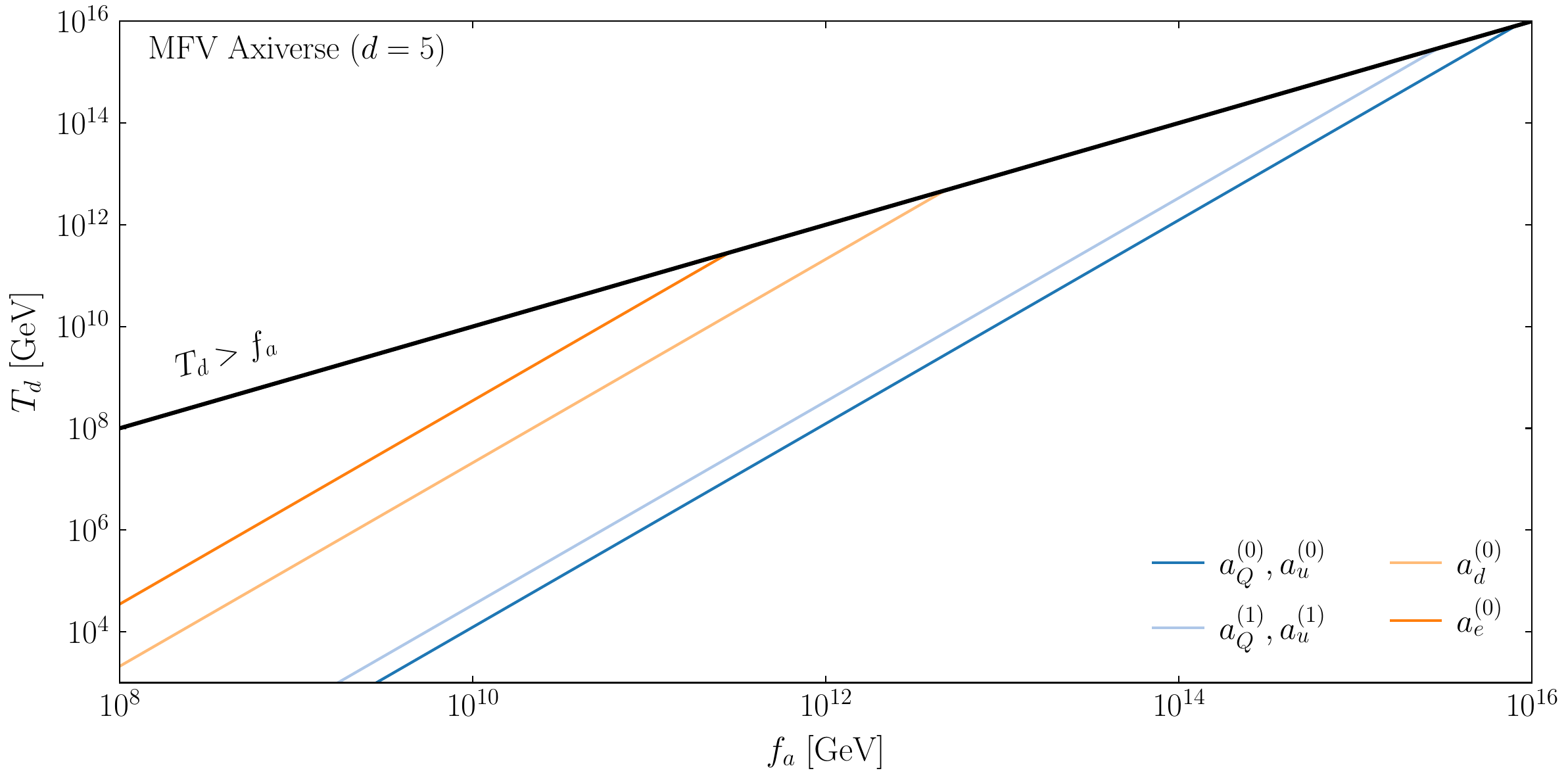}
    \caption{As in Fig.~\ref{fig:Hadronic_Tdec}, but assuming that the axion couplings to fermions satisfy MFV (Sec.~\ref{sec:MFV}). The axion states $a_F^{(i)}$ are defined as those which couple to fermions $F$ with strength $c_F^{(i)}$; see discussion after Eq.~\eqref{eq:MFV}.
    }
\label{fig:MFV_Tdec}
\end{figure}

This toy model includes exactly the relevant terms for the lepton-coupled axions, so $\Tilde{\Gamma}_{a_\mathsf{L}^{nm}} = \Tilde{\Gamma}_{a_L^{nm}}$ and $\Tilde{\Gamma}_{a_e^{nm}} = \Tilde{\Gamma}_{a_\mathsf{r}^{nm}}$. To apply it to the quark-coupled axions, note there is an additional factor of $N_c=3$ in the rate and that $a_Q^{nm}$ gets contributions from both the up and down Yukawas. We find that the CKM matrix $V$ always enters in the production rate as $\sum_j |V_{ij}|^2 = 1$ due to weak universality. Putting everything together, we have 
\ba{
\Tilde{\Gamma}_{a_Q^{nm}} &= |c_{nm}|^2\left((Y^u_{nn})^2 + (Y^u_{mm})^2 + (Y^d_{nn})^2 + (Y^d_{mm})^2\right)\dfrac{21\zeta(3)}{128\pi^3} \dfrac{T^3}{f_a^2}, \\
\Tilde{\Gamma}_{a_u^{nm}} &= |c_{nm}|^2\left((Y^u_{nn})^2 + (Y^u_{mm})^2\right)\dfrac{21\zeta(3)}{128\pi^3} \dfrac{T^3}{f_a^2}, \\
\Tilde{\Gamma}_{a_d^{nm}} &= |c_{nm}|^2\left((Y^d_{nn})^2 + (Y^d_{mm})^2\right)\dfrac{21\zeta(3)}{128\pi^3} \dfrac{T^3}{f_a^2}. \\
}
As stated above, we define $T_d$ as the (momentum-independent) solution to $\Tilde{\Gamma} = H$, which allows us to compute the resulting axion energy density as in Eq.~\eqref{eq:rho_dim5}.
The $T_d$ for various axion states are shown in Figs.~\ref{fig:Anarchy_Tdec},~\ref{fig:Texture_Tdec}, and~\ref{fig:MFV_Tdec}, for anarchy, texture, and MFV, respectively.

\subsubsection{Production via Dim-6 Higgs Coupling}\label{sec:dim6Higgs}
In~\eqref{eq:lamah_i}, we argued that the production of the $i$-th axion depends only on the coefficient $\lambda_{aH}^i$.
Thus, to determine this abundance, we use the following simplified interaction term,
\begin{equation}
\mathcal{L} \supset \dfrac{\lambda_{aH}}{f_a^2}(\partial_\mu a)^2 |H|^2.
\end{equation}
The production proceeds through Higgs-Higgs annihilation into two axions, which can be written 
\begin{equation}
H(\vecp_1,s_1) + \Hd(\vecp_2,s_2) \to a(\veck_1) + a(\veck_2).
\end{equation}
where $\vecp$ and $\veck$ are 3-momenta while $s_1,s_2\in\{1,2\}$ count internal degrees of freedom. The matrix element is simply given by 
\begin{equation}
    i\mathcal{M} = -\lambda_{aH}\dfrac{\mathsf{s}}{f_a^2}.
\end{equation}
neglecting axion masses.

As in the previous section, we wish to determine the axion distribution function $\f_a(\veck_1)$ by solving the Boltzmann equation with the collision term
\es{}{\label{eq:PSBE_H2}
C[\f_a] =& {1\over 2}\int \D \Pi_H \D \Pi_{\Hd} \D \Pi_a (2\pi)^4 \delta^4(k_1+k_2-p_1-p_2)|{\cal M}|^2\\ 
& \quad\quad\times (\f_H(\vecp_1,t) \f_{\Hd}(\vecp_2,t) - \f_a(\veck_1,t) \f_a(\veck_2,t)).
}
However, the fact that there are two axions in the final state leads to a nonlinear integro-differential equation which must be solved numerically. We leave this to future work, and make the approximation that we can ignore the backwards process: $\f_a(\veck_1,t) \f_a(\veck_2,t) = 0$. Clearly, this can only be justified in the regime that the axion does not thermalize; we ceiling $\Dneff$ at a thermal abundance. Following~\cite{Edsjo:1997bg,Hall:2009bx,Blennow:2013jba}, one can then solve the (integrated) Boltzmann equations for the axion number density,
\es{}{
\dfrac{\D n_a}{\D t}+3Hn_a &= \dfrac{T}{2^{11}\pi^6}\int_0^\infty \D \mathsf{s} \sqrt{\mathsf{s}} \,K_1\left(\sqrt{\mathsf{s}}/T\right) \sum_{\rm channels} \int \D \Omega\, |{\cal M}|^2 \\
&=  \dfrac{3\lambda_{aH}^2T^8}{\pi^5 f_a^4},
}
leading to a yield
\ba{
Y_a &= \dfrac{135 \lambda_{aH}^2}{2\pi^8} \sqrt{\dfrac{10}{g_{*}(\trh)g_{*s}^2(\trh)}}\dfrac{\mpl\trh^3}{f_a^4}.
}
While this result matches parametrically with Ref.~\cite{Bauer:2022rwf}, its numerical value differs by several orders of magnitude.
For a verification, we have checked that our computation matches with 
micrOMEGAs~\cite{Alguero:2023zol} numerical output.

In the above, we have assumed there is only a single axion.
To compute an energy density, we require the average energy density of each axion, which is unknown. We assume that the axions have a thermal distribution with $\langle E \rangle = (\pi^4/30)\zeta(3)\, T$, although the fact that the cross section $\sigma(\mathsf{s}) \propto \mathsf{s}$ means that we likely are underestimating the true value. Furthermore, we must account for SM entropy depositions, which dilute the axion energy density by an additional factor $(g_{*s}(T_0)/g_{*s}(\trh))^{4/3}$. This leads to
\ba{
\Delta N_{\rm eff} \approx \min\left(0.027, 310\,\lambda_{aH}^2 \left(\dfrac{10^{12}\ {\rm GeV}}{f_a}\right)^4 \left(\dfrac{T_{\rm RH}}{10^{12}\ {\rm GeV}}\right)^3.
\right)}
Here, the min function ensures that when the axion obtains a thermal abundance, the backwards process becomes important and $\Dneff$ reaches the value of $0.027$, as appropriate for a single thermalized scalar degree of freedom.

\subsubsection{Production via Dim-6 Charge Radius Operator}

We now consider production of axions due to the antisymmetric dimension-6 charge-radius operator
\begin{equation}
\mathcal{L} \supset \sum_{F,i,j} \dfrac{\mathcal{Y}_F c_{aB}^{ij}}{f_a^2} \phi_{[i}\partial_\nu \phi_{j]} \bar{\Psi}_F \gamma^\nu \Psi_F
.
\end{equation}
The production proceeds through fermion annihilation into two distinct axions, which can be written 
\begin{equation}
\Psi(\vecp_1,s_1) + \Psibar(\vecp_2,s_2) \to \phi_i(\veck_1) + \phi_j(\veck_2).
\end{equation}
We note there are two other production processes with $\phi_j$ in the initial state, but those are subdominant if we assume that $\phi_j$ has zero initial abundance. The squared matrix element is given by 
\begin{equation}
    |\mathcal{M}|^2 = -4 (c_{aB}^{ij})^2 \mathcal{Y}_F^2 \dfrac{\mathsf{t} (\mathsf{s}+\mathsf{t})}{f_a^4}.
\end{equation}
We would like to compute the axion distribution function $\f_i(\veck_1)$ by solving the Boltzmann equation with collision term
\es{}{\label{eq:PSBE_B}
C[\f_i] =& {1\over 2}\int \D \Pi_\Psi \D \Pi_{\Psibar} \D \Pi_j (2\pi)^4 \delta^4(k_1+k_2-p_1-p_2)|{\cal M}|^2 \\
& \quad \quad \times (\f_\Psi(\vecp_1,t) \f_{\Psibar}(\vecp_2,t) - \f_i(\veck_1,t) \f_j(\veck_2,t)).
}
However, $\f_j$ is entirely unknown and so we must solve the coupled system of Boltzmann equations over all $i$. 
Similar to above, we simplify the calculation by neglecting the backwards process and solving the Boltzmann equation for the $\phi_i$ number densities.
Using the notation~\eqref{eq:lamab_i}, we find
\ba{
\Delta N_{\rm eff} \approx \min\left(0.027,220\, (\lambda_{aB})^2 \left(\dfrac{10^{12}\ {\rm GeV}}{f_a}\right)^4 \left(\dfrac{T_{\rm RH}}{10^{12}\ {\rm GeV}}\right)^3
\right)}
under the same assumptions as for the symmetric dimension-6 operator. Here we have summed over all possible initial particles and computed $\sum |{\cal Y}_F|^2$ as appropriate for the SM particles. 

For each axion state uncoupled at dimension five and for $\lambda_{aB}=1=\lambda_{aH}$, we then have
\ba{
\label{eq:dim6_neff}
\Delta N_{\rm eff} \approx \min\left(0.027,530\,\left(\dfrac{10^{12}\ {\rm GeV}}{f_a}\right)^4 \left(\dfrac{T_{\rm RH}}{10^{12}\ {\rm GeV}}\right)^3
\right).}
We crudely estimate a decoupling temperature for the collective dimension-6 processes by equating the two arguments of the min function above:
\ba{
T_d^{\rm dim-6} \approx 3.7 \times 10^{10} \left(\dfrac{f_a}{10^{12}\ {\rm GeV}}\right)^{4/3}\,{\rm GeV}.}

\section{Results for Benchmark EFTs}
\label{sec:result}
\begin{figure}[!htb]
    \centering
    \includegraphics[width=\textwidth]{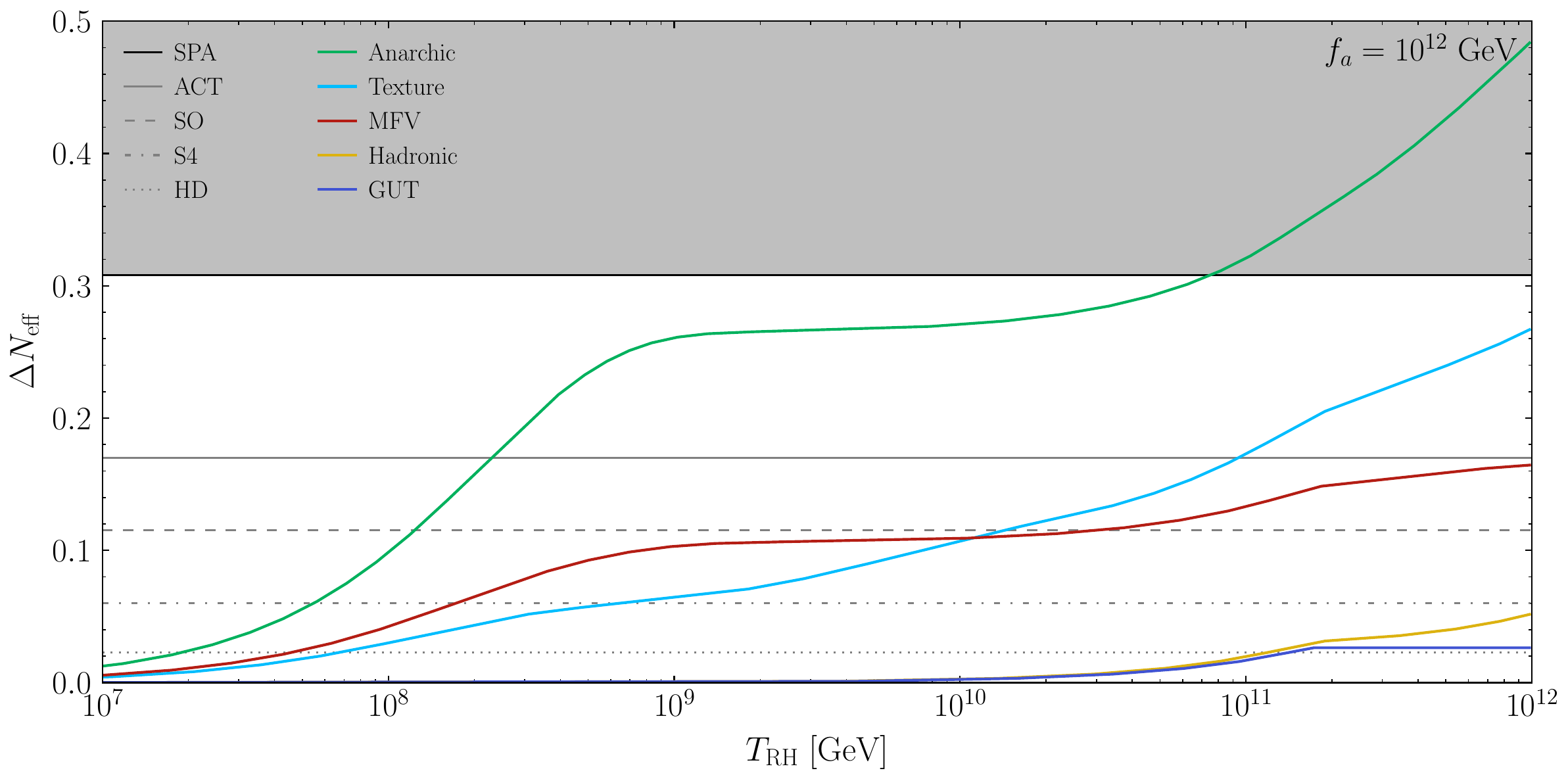}
    \caption{A summary of our results considering only dimension-5 couplings for $f_a = 10^{12}$ GeV as a function of $\trh$ and different assumptions about the axion couplings. The existing and future CMB constraints are shown as horizontal lines: the combination of SPT-3G, Planck, and ACT (SPA, black solid), ACT alone (gray solid), SO (dashed), S4 (dot-dashed), and HD (dotted). The $\Dneff$ expected are also shown assuming {anarchic} couplings, a {Froggatt-Nielsen texture}, {MFV}, {hadronic}, and a {hadronic GUT}. The kink on the hadronic GUT (blue) line around $2\times 10^{11}$~GeV indicates the decoupling temperature $T_d$ of the GUT axion (for $f_a = 10^{12}$~GeV). For $\trh > T_d$, the GUT axion is thermalized with the SM bath and thus $\Dneff$ no longer rises with $\trh$.
    }
\label{fig:dim5_1D}
\end{figure}

In this section, we present the predicted value of $\Dneff$ under the various scenarios for axion couplings outlined in Sec.~\ref{sec:benchmark}.
Before discussing the individual cases in more detail, in Fig.~\ref{fig:dim5_1D} we provide a summary of how $\Dneff$ varies with the reheat temperature $\trh$ for dimension-5 couplings and $f_a=10^{12}$~GeV\@. 
As expected, the predicted value of $\Dneff$ scales with the number of accessible axion states, with anarchic couplings producing the most radiation and the GUT case the least. 
In this plot, we assume that the number of axion states is larger than the number of SM interactions ($\N\geq 44$), and we do not include the effects of dimension-6 interactions.
Notably, there is rich variation in $\Dneff$ as a function of $\trh$ among different benchmarks. For example, texture allows for a larger number of independent axion couplings with a large enough coupling (30, from Fig.~\ref{fig:Texture_Tdec}) compared to MFV (6, from Fig.~\ref{fig:MFV_Tdec}). One might thus expect that $\Dneff$ for texture would also be larger compared to MFV. However, $\Dneff$ depends on the individual $T_d$ for each axion, which vary non-trivially among different benchmarks. As a result, for $\trh \sim 10^8-10^{10}$~GeV and $f_a=10^{12}$~GeV, a larger number of axions are thermally produced in MFV, compared to texture, leading to a larger $\Dneff$. This is consistent with the results in Figs.~\ref{fig:Texture_Tdec} and~\ref{fig:MFV_Tdec}.

We also show the sensitivity of current and upcoming CMB experiments. The horizontal lines correspond to existing one-sided 95\% upper limits $\Dneff^{95\%}$. 
SPT-Planck-ACT (SPA) corresponds to the constraint from the recent SPT-3G analysis~\cite{SPT-3G:2025bzu} when floating the helium abundance $Y_P$ in the fit, yielding $N_{\rm eff}=2.99_{-0.26}^{+0.22}$, so that $\Dneff^{95\%}\approx0.31$. We note that without floating $Y_P$, we find $\Dneff^{95\%}\approx-0.04$ due to the low central values from Planck and ACT\@. For this reason, we adopt the more conservative result from floating $Y_P$.
Indeed, the ACT results~\cite{ACT:2025tim} (including Planck lensing and BOSS BAO) yield $\Dneff^{95\%}\approx0.04$. We show instead the result from restricting to physical parameter space with $\Dneff>0$: $\Dneff^{95\%}=0.17$, but in plots we shade only the region excluded by the SPA analysis. Finally, we show projections for Simons Observatory (SO), CMB-S4 (S4), and CMB-HD (HD)~\cite{SimonsObservatory:2025wwn,CMB-S4:2016ple,MacInnis:2023vif}.

\begin{figure}[!htb]
    \centering
    \includegraphics[width=\textwidth]{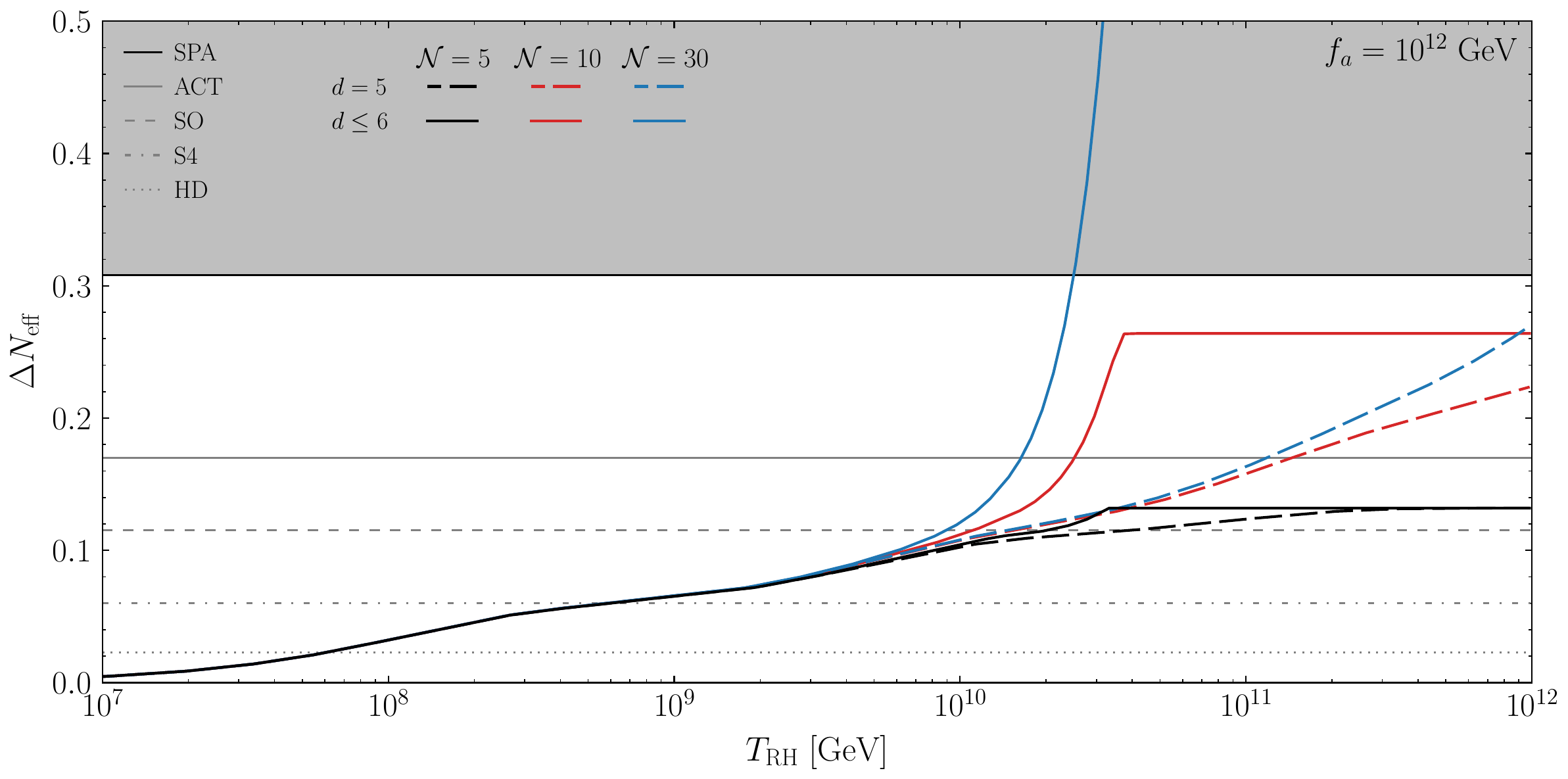}
    \caption{A comparison of our dimension-5 and dimension-6 results for the Froggatt-Nielsen textured case and for $f_a = 10^{12}$ GeV as a function of $\trh$ and $\mathcal{N}$. The existing and future CMB constraints are shown as in Fig.~\ref{fig:dim5_1D}. The $\Dneff$ expected are shown with (dashed) and without (solid) $D=6$ operators on for {$\mathcal{N}=5$}, {$\mathcal{N}=10$}, and {$\mathcal{N}=30$}. The kinks on the solid lines indicate the decoupling temperature $T_d \sim 3 \times 10^{10}$~GeV of the dimension-6 operator (for $f_a = 10^{12}$~GeV). For $\trh > T_d$, all the axions are thermalized and $\Dneff$ is saturated.
    }
\label{fig:1D_Texture_Comparison}
\end{figure}

In Fig.~\ref{fig:1D_Texture_Comparison}, for the assumption of textured couplings, we illustrate the joint effects of assuming different $\N$ and turning on dimension-6 couplings. For this value of $\fa$, the decoupling temperature associated with the $d=6$ interactions is $T_d\sim3\times10^{10}$ GeV\@. 
At that temperature, only four dimension-5 interactions would thermalize their respective axion states. The remaining states are typically thermalized through $d=6$ interactions due to the stronger $T-$dependence of the production rates.

In the remainder of this section we show the results for $\Dneff$ for each of the benchmark EFTs as a function of $\trh$ and $f_a$.

\subsection{Hadronic Axiverse}

We first consider the hadronic axiverse (Eq.~\eqref{eq:had_ax}), where the axions couple to the SM only through gauge interactions. Because there are only three such axion-SM operators, we see immediately that we will have at most $\Dneff \sim 3\times0.027$ (for decoupling temperature above the electroweak scale). We expect that the three axions will have decoupling temperatures scaling as $T_d \sim 1/\alpha_i^3$, so that the QCD axion $a_{\rm QCD}$ decouples at the lowest temperature and the hypercharge axion $a_1$ decouples at the highest temperature, at least below the unification scale. This is what we observe in Fig.~\ref{fig:Hadronic_Tdec}, where we plot the decoupling temperatures of the individual axion states as a function of $f_a$. We have made use of the axion-gauge interaction rates computed at finite temperature in Ref.~\cite{Salvio:2013iaa}, and assumed the instantaneous decoupling approximation. The space above the solid black line, labeled $T_d > f_a$, corresponds to the parameter region where the EFT breaks down. 
We do not make any claims about temperatures larger than $f_a$ because new states may be accessible at these energies that could drastically change our predictions. 
As a simple example, in the single-axion KSVZ model, the heavy quarks could thermalize with the bath and later decay to axions (see, {\it e.g.}, Ref.~\cite{Cheek:2023fht}). 
In principle, such additional states could even appear at scales lower than $f_a$, so that our EFT breaks down earlier; however, we assume that our EFT is valid up to $f_a$.

\begin{figure}[!htb]
    \centering
    \includegraphics[width=\textwidth]{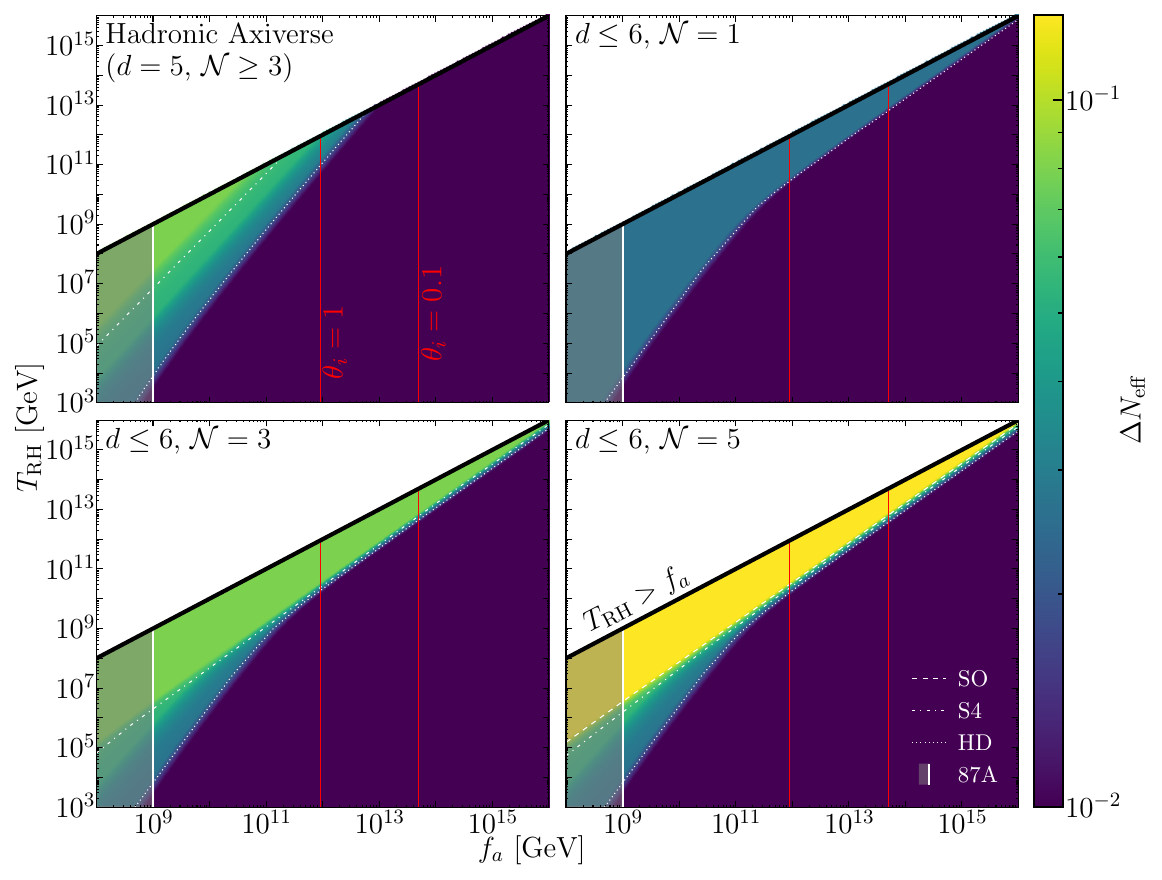}
    \caption{{\it (Upper Left)} The axion contribution to $\Dneff$ as a function of $\trh$ and $f_a$, assuming that the axion couplings to the SM are hadronic~(Eq.~\eqref{eq:had_ax}), that there are at least three axion states $\N\geq 3$, and not accounting for contributions from the dimension-6 operators. The white space indicates the parameter space where the EFT breaks down $\trh > f_a$. The dashed white contour lines indicate the sensitivity of future CMB surveys SO, S4, and HD, to this scenario. The shaded regions correspond to excluded regions, in this case due to limits from SN 1987A (labeled `87A'). The red vertical lines denote the $f_a$ at which the QCD axion makes up all of the DM for the specified initial misalignment angle $\theta_i$, and apply to all panels. {\it (Upper Right)} Now accounting for dimension-6 operators, and assuming $\N=1$. {\it (Lower Left)} As in the upper right panel, but assuming $\N=3$. {\it (Lower Right)} As in the upper right panel, but assuming $\N=5$.
    }
\label{fig:Hadronic_Density_2x2}
\end{figure}

In Fig.~\ref{fig:Hadronic_Density_2x2}, we show the $\Dneff$ predictions in the hadronic axiverse as a function of the reheating temperature $\trh$ and $\fa$. Observations of NS cooling~\cite{Buschmann:2021juv} and SN 1987A~\cite{1990PhR...198....1R,Springmann:2024mjp} impose a lower limit of $f_a \geq 3.6 \times 10^8$ GeV and $f_a \geq 10^9$ GeV, respectively. In the upper left, we show results with $d=6$ operators turned off, assuming that $\N\geq3$ ({\it i.e.}~the number of axion-SM operators). This does not correspond necessarily to a realistic theory, but is meant to be indicative of the $d=5$ effects. The purple area corresponds to the region where the reheat temperature is too low and dark radiation production is negligible. 
As one moves in the plot to the upper left, axion production increases. The blue region corresponds to the region in which the QCD axion thermalizes with the SM bath ($\trh>T_{d,{\rm QCD}}$, and the last two green regions correspond to those where the $a_2$ and $a_1$ axions thermalize, respectively (see Fig.~\ref{fig:Hadronic_Tdec} for a direct comparison).
CMB-S4 is required to probe any part of this parameter space, whereas CMB-HD is needed to probe down to the QCD axion decoupling line.
For $f_a<10^{12}$~GeV, these telescopes would be sensitive to cosmological scenarios with parametrically small values of $\trh/\fa$. 
On the other hand, for $\fa \gtrsim 10^{12}$ GeV, the QCD axion may contribute to the DM abundance via the misalignment mechanism~\eqref{eq:misal}.
To show this, in Fig.~\ref{fig:Hadronic_Density_2x2} we indicate the values of $f_a$ such that the QCD axion can explain the DM abundance for two choices of the initial misalignment angle $\theta_i=0.1,1$.

In the upper right panel of Fig.~\ref{fig:Hadronic_Density_2x2}, we show the expected $\Dneff$ including dimension-6 interactions for a single-axion theory. At low $\fa$, the production is dominated by the QCD interactions, while at higher $f_a>{\rm few}\times10^{11}$ GeV the $d=6$ interactions dominate. 
In the lower left panel of Fig.~\ref{fig:Hadronic_Density_2x2}, we show the same but for $\N=3$. For $\trh$ within a couple orders of magnitude of $\fa$, the $d=6$ interactions thermalize all axion states, which extends the reach of CMB-S4. This effect continues in the lower right panel, where $\N=5$ is enough to observe a signal at SO\@.
Overall, this highlights the importance of the dimension-6 operators and the {\it total} number of axions $\N$ in the axiverse.

\begin{figure}[!htb]
    \centering
    \includegraphics[width=\textwidth]{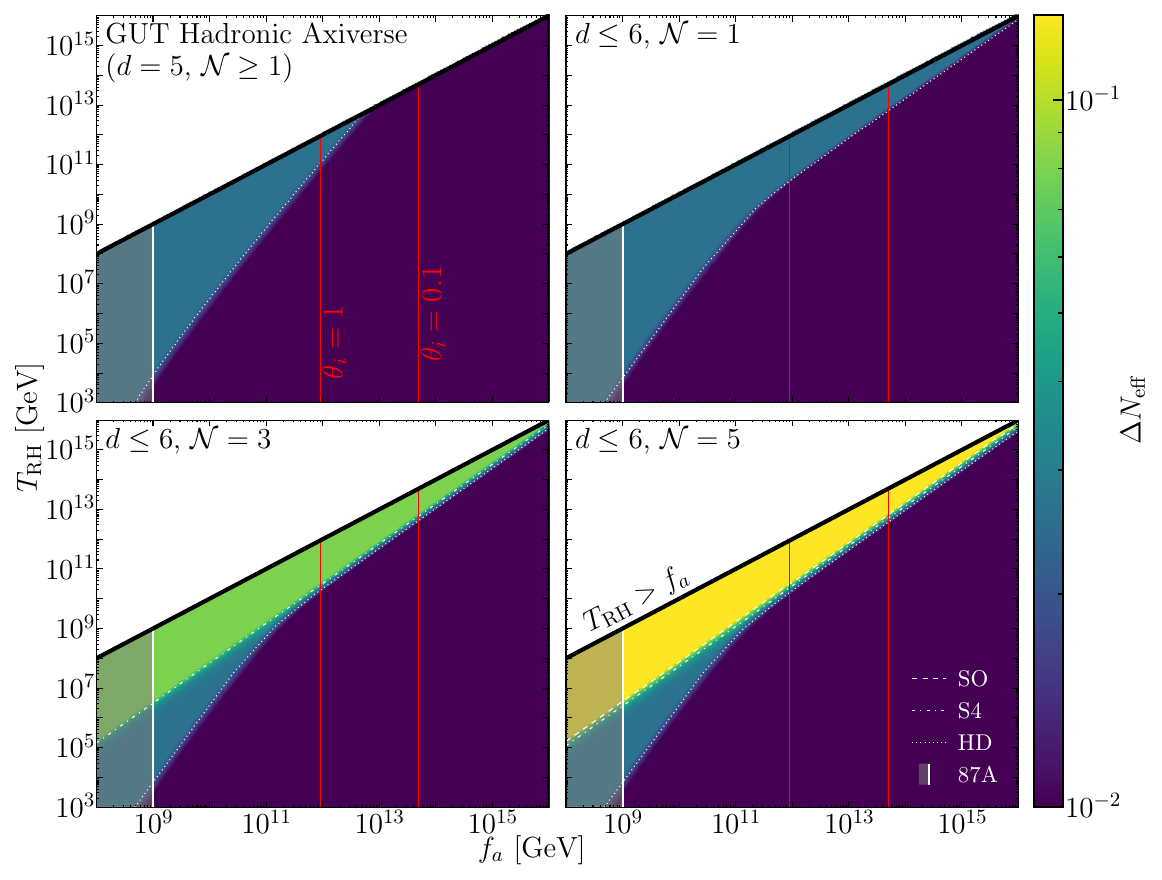}
    \caption{As in Fig.~\ref{fig:Hadronic_Density_2x2}, but with the additional assumption that the SM unifies into a GUT, so that all three axions coupled to gauge bosons are aligned~(Eq.~\eqref{eq:gut_ax}). Note that the upper panels are identical to that for a single QCD axion. 
    }
\label{fig:GUT_Density_2x2}
\end{figure}

If the SM unifies into a simple GUT, then the axions coupled to each gauge group are aligned~\cite{Agrawal:2022lsp}, although this is not necessarily realized in semi-simple GUTs, such as Pati-Salam or Trinification. 
Thus, in simple GUTs only one axion interacts with the SM~(Eq.~\eqref{eq:gut_ax}), and its interactions at energies below the GUT scale are dominated by QCD\@. 
This is why the decoupling temperature for $a_{\rm GUT}$ is nearly identical to the decoupling temperature for the QCD axion in Fig.~\ref{fig:Hadronic_Tdec}. 
We see in the upper left panel of Fig.~\ref{fig:GUT_Density_2x2} that only one axion can thermalize, leading to a smaller $\Dneff = 0.027$, which CMB-HD would be required to observe. 
In the remaining panels of Fig.~\ref{fig:GUT_Density_2x2}, we include $d=6$ operators and consider different values of $\N$.
We observe a similar pattern as Fig.~\ref{fig:Hadronic_Density_2x2}, with sensitivity to both $d=6$ interactions and the value of $\N$.

\subsection{Fermionic Axiverse}
In this section we consider axiverse couplings which are fermionic, which we take to mean that the axion states couple to SM fermions in addition to gauge bosons. {\it A priori}, the strength and the flavor structure of the couplings to fermions are unknown, so we show results for the anarchic, Froggatt-Nielsen, and MFV cases discussed in Sec.~\ref{sec:benchmark}.

\subsubsection{Anarchic Axiverse}

\begin{figure}[!htb]
    \centering
    \includegraphics[width=\textwidth]{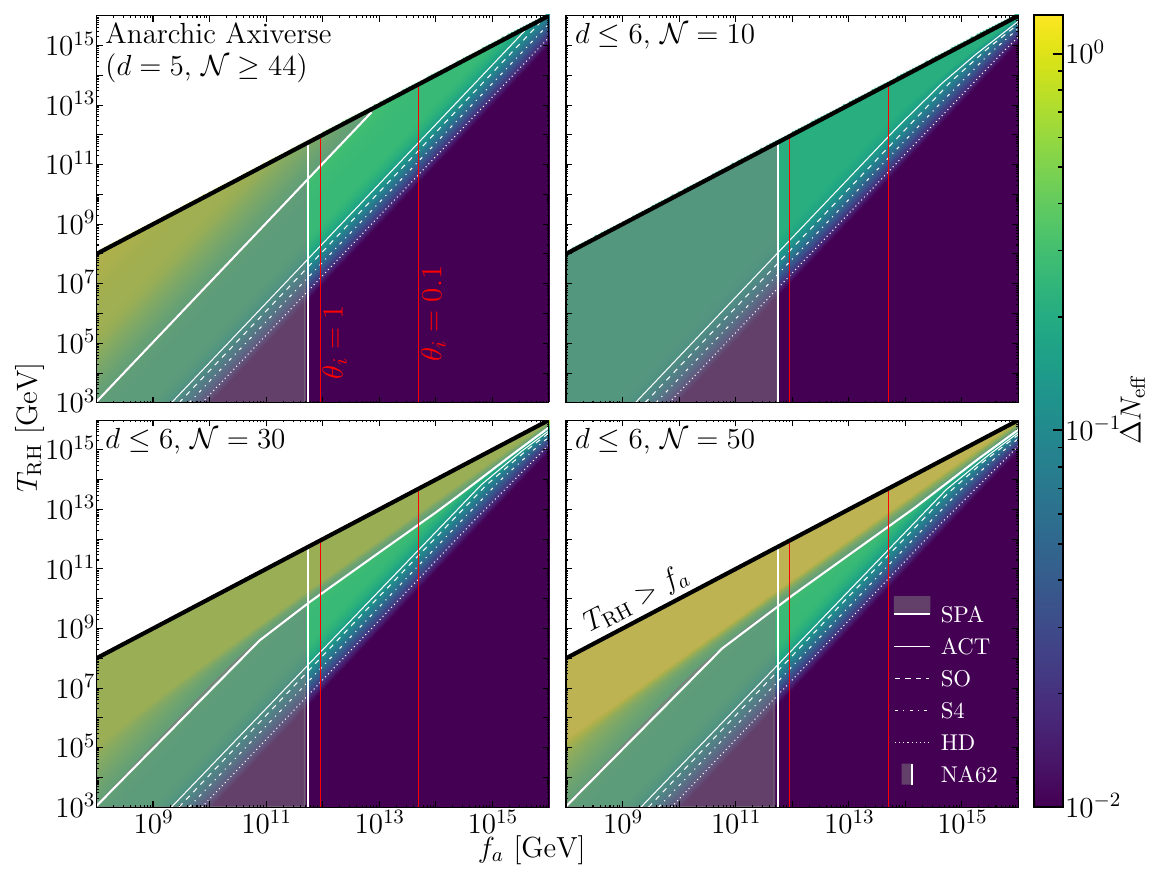}
    \caption{As in Fig.~\ref{fig:Hadronic_Density_2x2}, but assuming the axion couplings to the SM fermions are anarchic. Searches for flavor violation at NA62 constrain $\fa/c_{Q,12}, \fa/c_{d,12}\gtrsim 6\times 10^{11}$ GeV\@.
    The shaded region is excluded by a combination of this and overproduction of $\Dneff$ as measured by SPA\@. We show current sensitivity from ACT alone and future sensitivity for SO, S4, and HD\@. Note that the upper left panel assumes $\N\geq44$, and the other three panels assume $\N=10,30$, and $50$, respectively.
    }
\label{fig:Anarchy_Density_2x2}
\end{figure}

Anarchic axion-fermion couplings~(sec.~\ref{sec:anarchy}) lead to the largest possible $\Dneff$ contributions, as each coupling $c_{F,nm}$ in Eq.~\eqref{eq:Lagrangian} takes on an unsuppressed $\mathcal{O}(1)$ value. We show this in Fig.~\ref{fig:Anarchy_Tdec}, where we show the decoupling temperature $T_d$ of each axion state which is coupled to fermions (the axion-gauge couplings lead to $T_d$ as in the hadronic case, and are not shown). As expected, the axions which couple to the top (even in a flavor-violating way) have the lowest $T_d$, which scales like $1/y^{2}$, for a Yukawa coupling $y$. There are ten such axion states: the five coupled to the third generation of $Q$, and the five coupled to the third generation of $u$, in total eight of which are flavor-violating. 
As can be seen in the upper left panel of Fig.~\ref{fig:Anarchy_Density_2x2}, if $\trh$ is large enough, all ten axions are thermalized at early times.
Such a scenario is excluded by the recent ACT data, even for $\fa$ close to the GUT scale $\sim 10^{16}$~GeV, although SPA only excludes parameter space for which additional axions are thermalized. 
Future CMB surveys will improve the current sensitivity by several orders of magnitude. Due to the large flavor-violating couplings in the anarchic axiverse, beam-dump experiments strongly constrain even light axions. The strongest beam-dump constraint comes from NA62 searches for $K^+ \to \pi^++{\rm invisible}$~\cite{NA62:2024pjp}. 
Ref.~\cite{Guadagnoli:2025xnt} finds $\fa/c_{Q,12}, \fa/c_{d,12}\gtrsim 6\times 10^{11}$ GeV, far stronger than the bound from NS cooling. We shade the region excluded by flavor-violation searches in Fig.~\ref{fig:Anarchy_Density_2x2}. 
Additionally, for $\theta_i\sim {\cal O}(1)$, the overclosure bound from QCD axion DM production~\eqref{eq:misal} requires $f_a < 9 \times 10^{11}$~GeV\@.
This choice of $\theta_i$, along with the above-mentioned Kaon decay bound with $c_{Q,12}, c_{d,12}\sim 1$, rules out most of the $\trh-f_a$ parameter space, except for the narrow region bounded by the NA62, SPA, and the $\theta_i=1$ lines.
As expected, if one allows for a moderate fine-tuning of $\theta_i=0.1$, the parameter space to the left of the $\theta_i=0.1$ line opens up.
Similarly, a smaller $c_{Q,12}, c_{d,12}$ would move the flavor bounds towards the left, opening up new parameter space as well.
In the additional three panels, we show results including dimension-6 operators for $\N=10$, 30, and 50. As expected, for sufficiently large $\N$, values of $\trh$ near $\fa$ are excluded by existing CMB data.

\subsubsection{Froggatt-Nielsen Textured Axiverse}

The assumption that the Froggatt-Nielsen mechanism solves the flavor puzzle leads to a suppression of flavor-violating axion couplings relative to the anarchic case, by a factor of $\epsilon$ raised to some power. Unsurprisingly, then, the decoupling temperatures of the flavor-violating axion states increase, so that such a scenario predicts a smaller $\Dneff$ than the anarchic case discussed above. We see this behavior in Fig.~\ref{fig:Texture_Tdec}. Explicitly, the two axion states coupled to $t_R$ and $c_R$ (labeled $a_u^{32}$ and $a_u^{23}$) have $T_d$ larger by a factor $\epsilon^{-2}$ relative to anarchy (note, $\epsilon\approx 0.23$), while those coupled to $t_L$ and $c_L$ (labeled $a_Q^{32}$ and $a_Q^{23}$) have $T_d$ larger by a factor $\epsilon^4$. This can be verified by plugging in the explicit charges chosen in Eq.~\eqref{eq:texture}.

\begin{figure}[!htb]
    \centering
    \includegraphics[width=\textwidth]{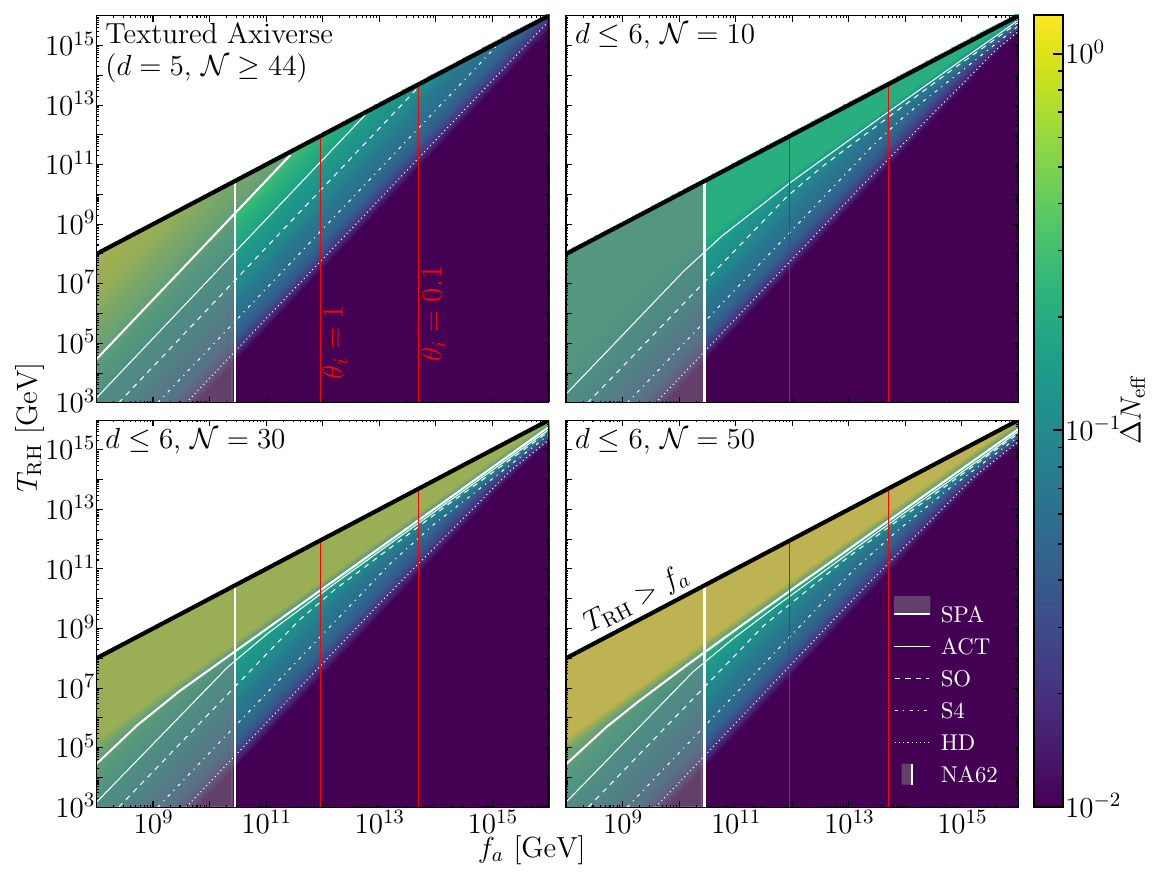}
    \caption{As in Fig.~\ref{fig:Anarchy_Density_2x2}, but assuming the axion couplings to the SM fermions arise from a Froggatt-Nielsen texture~(Eq.~\eqref{eq:texture}). This weakens the NA62 constraint on flavor violation to $f_a\gtrsim2\times10^{10}$ GeV\@.
    }
\label{fig:Textured_Density_2x2}
\end{figure}

Because of this suppression, the $\Dneff$ constraints are significantly relaxed in Fig.~\ref{fig:Textured_Density_2x2}. On the other hand, the NA62 constraint on $K^+\to\pi^+a$ is only suppressed by a factor $\epsilon^2$, so that it is still much stronger than NS cooling. We find that SPA excludes part of the parameter space up to \journal{$\fa \sim 10^{12}$~GeV} even if dimension-6 operators are not active.
Future CMB experiments can improve constraints by as much as four orders of magnitude in some parts of the parameter space.

\subsubsection{MFV Axiverse}

\begin{figure}[!htb]
    \centering
    \includegraphics[width=\textwidth]{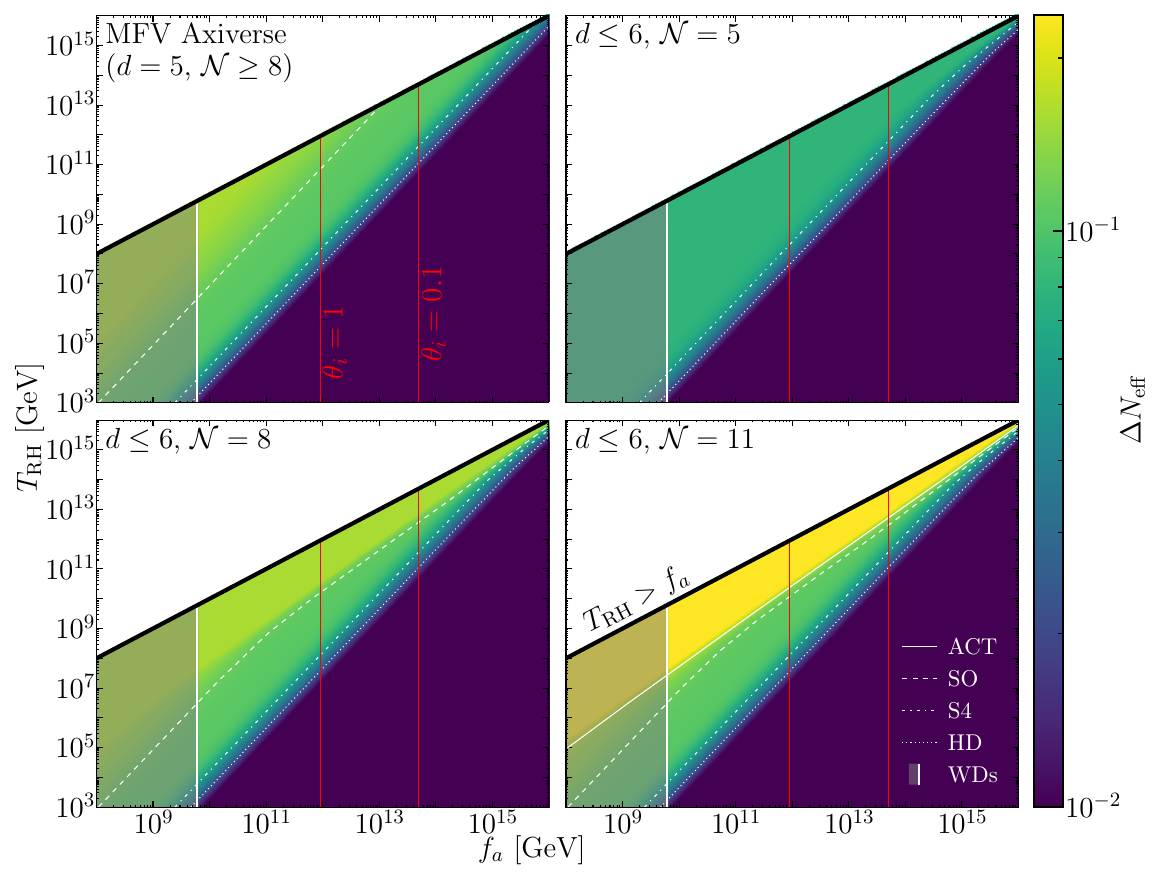}
    \caption{As in Fig.~\ref{fig:Hadronic_Density_2x2}, but assuming the axion couplings to the SM fermions satisfy MFV\@. Observations of WD cooling, arising from the axion-electron coupling, constrain $f_a/|c_{ee}|\gtrsim6.1\times10^9$\,GeV, and we show the bound with $|c_{ee}|=1$ in the figure (note $c^L_{11}=0$ in our basis, so $c_{ee}=c^e_{11}$). The upper left panel requires $\N\geq8$, while the remaining panels show $\N=5,8,$ and $11$, respectively.
    }
\label{fig:MFV_Density_2x2}
\end{figure}

We finally investigate a scenario in which the axion couplings are described by MFV\@. 
As we saw in Sec.~\ref{sec:MFV}, only six axion states are non-trivially coupled to the SM fermions. 
By construction, flavor violation is extremely small, and beam-dump constraints are far subdominant to flavor-conserving constraints. For $\mathcal{O}(1)$ coupling, the most stringent limit on $\fa$ comes from white dwarf (WD) cooling~\cite{2015ApJ...809..141H}, requiring $\fa\gtrsim 6.1\times10^9$ GeV\@, which of course becomes weaker if the dimensionless couplings are smaller.
We see in Fig.~\ref{fig:MFV_Tdec} that the four states which are coupled to $Q$ and $u$ have low $T_d$ through their interactions with the top quark, while the two states coupled to $d$ and $e$ are suppressed by the smaller $y_b$ and $y_\tau$, respectively. In the upper left panel of Fig.~\ref{fig:MFV_Density_2x2}, we show the resulting $\Dneff$ for $d=5$ operators only, which would require SO to probe. In the remaining three panels, we show $\N=5$, 8, and 11 where the $d=6$ contributions get successively larger. 

\subsection{Contributions from Dimension-6 Operators}
\begin{figure}[!htb]
    \centering
    \includegraphics[width=\textwidth]{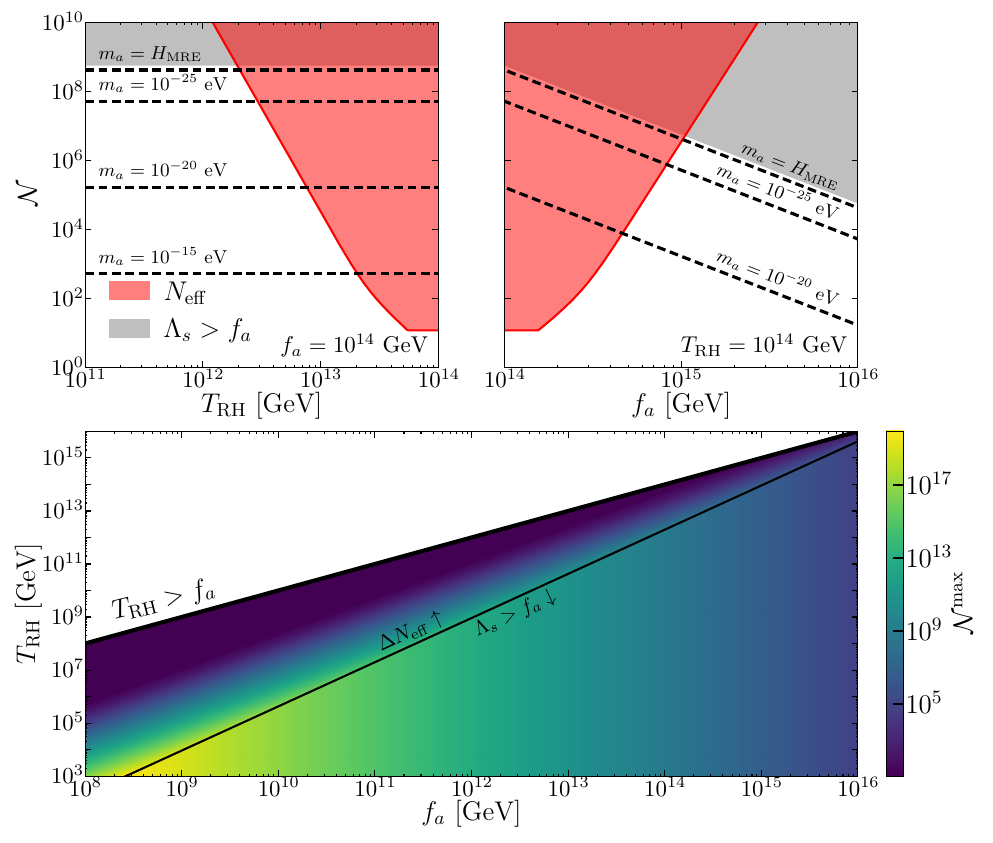}
    \caption{{\it (Upper Left)} Upper limits on $\N$ from the species scale (gray) and from $\Dneff$ (red), as a function of $\trh$, for $f_a=10^{14}$ GeV\@. \journal{The dashed black lines indicate the axion mass such that, if all axions have the same mass $m_a$ and initial misalignment angle $\theta_i = 1$, gives the correct DM abundance via misalignment. This assumes there is no misalignment contribution from QCD, which generically overproduces DM at the shown $f_a$. For the smallest ${\cal N}=12$ constrained by $N_{\rm eff}$, one requires $m_a \approx 2 \times 10^{-12}\, (f_a/10^{14}\ {\rm GeV})$ eV.} {\it (Upper Right)} As in the upper left, but as a function of $f_a$ for $\trh=10^{14}$ GeV\@. {\it (Lower Panel)} As in the upper panels, but over the full 2D parameter space. The colors correspond to the maximum $\N$ allowed by either the species scale or $\Dneff$, as labeled.
    }
\label{fig:max_N}
\end{figure}

As the dimension-6 operators generically couple every axion to the SM, $N_{\rm eff}$ provides an upper limit on $\N$. We find that this upper limit can be much more stringent than that by requiring the species scale~\cite{Dvali:2007hz} $\Lambda_s>f_a$, corresponding to $\N \leq (M_{\rm Pl}/f_a)^2$ (up to a model-dependent $\mathcal{O}(1)$ factor we have ignored). The contribution to axion energy density from dimension-6 operators scales like
\ba{
\label{eq:dim6_neff_tot}
\Delta N_{\rm eff} \approx \min\left(0.027,530\,\left(\dfrac{10^{12}\ {\rm GeV}}{f_a}\right)^4 \left(\dfrac{T_{\rm RH}}{10^{12}\ {\rm GeV}}\right)^3
\right)\times\ {\N \over \sqrt{1+\N/106.75}},}
where in the last factor the numerator comes from the $\N$ distinct states and the denominator arises from the increase in $g_{*s}$. In the upper left (right) panel of Fig.~\ref{fig:max_N}, we show limits on $\N$ from the dimension-6 production in this work and from the species scale for fixed $\fa\ (\trh) = 10^{14}$ GeV\@. The $d=6$ operators can be much more stringent for large $\trh$ or low $\fa$. In the lower panel, we show the same as a 2D parameter space in $\trh$ and $\fa$, where $\N^{\rm max}$ is the minimum of the two limits.

\section{Conclusions}
\label{sec:conc}
The strong CP problem and the axion quality problem motivate an extra-dimensional origin of the QCD axion, and an associated axiverse, which predicts many light pseudoscalar degrees of freedom in the 4D EFT\@. 
On the other hand, the CMB and BBN measurements of $\Dneff$ are extremely strong probes of light, weakly coupled particles, like axions, that are nevertheless efficiently produced in the hot, dense primordial plasma of the early universe. Only up to $\mathcal{O}(10)$ axion states can thermalize with the bath at any point in the cosmological history, at odds with $\mathcal{O}(100)$ or more axions predicted in many concrete realizations of extra-dimensional topologies. For the first time, we write down the EFT up to dimension six for an arbitrary number of axions interacting with the SM\@. Other than the QCD axion, we assume all the axions have masses $\lesssim$~eV, so that they behave as `dark radiation' during CMB decoupling.

We find there are a maximum of 44 independent SM couplings at dimension five, so that only 44 independent axion states couple to the SM\textemdash although not all interact strongly enough to thermalize, even for the highest reheating temperatures and for flavor-anarchic couplings to fermions. 
We characterize the constraints and detection prospects for scenarios where axion-fermion couplings are determined by anarchy, Froggatt-Nielsen textures, minimal flavor violation, and where fermion couplings are absent at tree-level (like in KSVZ-type scenarios). 
Axion-fermion couplings may be organized by some other principle; to our knowledge no calculation of those couplings has been performed in explicit string constructions as in~\cite{Gendler:2023kjt} for the axion-photon coupling. Such a calculation would be very valuable in narrowing down the expected freeze-in abundance of string theory axions. The dimension-6 operators, on the other hand, are quadratic in the axion fields and can freeze-in the full axion spectrum. There too, no top-down calculation exists, and in our EFT treatment we assume $\mathcal{O}(1)$ Wilson coefficients. 

At the same time, in this work we have considered only the SM plus massless axions with reheating above the electroweak scale in the context of standard radiation-dominated cosmology. It would be interesting to explore departures from this setup. For example, if there are additional new fields, there would generically be additional operators. With the addition of a SM singlet scalar $\chi$, one could write down $\chi(\partial a)^2$ which couples all $\N$ axions at dimension five. In general, adding beyond-the-SM fields will increase $\Dneff$ beyond what we have found here. Incorporating heavy axions ($m_a \gtrsim$ eV) may also lead to interesting effects on our results. Beyond their contribution to DM through misalignment, mass mixing between axions may be an important feature of these models. Furthermore, if they make up an important component of DM, they would affect our assumption of standard cosmology if they, {\it e.g.}, lead to early-matter dominated eras or provide entropy injections through their decays. In any of these extensions, by considering $\trh$ to be the largest temperature below which standard radiation-dominated cosmology occurs, our work provides a conservative lower bound on the $\Dneff$ expected from the axiverse. One may also consider low-scale reheating at or below the weak scale. While below the weak scale, at most 44 axions may contribute, due to increasing $g_*$, $\Dneff$ can be larger than that considered here.

In computing $\Dneff$ from the gauge sector and from the dimension six operators, we have used a simplified framework based on the Boltzmann equations for the axion number density that sets the abundance to its thermal value above $T_d$, and scales it down by appropriate powers of $\Gamma/H|_{T=T_{\rm RH}}$ below $T_d$. This approach leads to $\mathcal{O}(1)$ errors in the vicinity of $T_d$. We leave a numerical evaluation of the phase-space density Boltzmann equations~\eqref{eq:PSBE_H2} and \eqref{eq:PSBE_B} for future work. We additionally ignored thermal corrections to the interaction rates (except in the gauge sector), although we note that our kinetic computations of the interaction rates due to flavor-conserving fermion couplings agrees with those in Ref.~\cite{Salvio:2013iaa} in the context of thermal field theory. 

Overall, using the current CMB surveys, we find a wide variety of high-scale reheating scenarios are inconsistent with $\Dneff$ constraints if there are a large number of axions in Nature.
The ongoing Simons Observatory or future CMB surveys will significantly improve the sensitivity and may reveal the first signatures of a cosmological axion background.

\section*{Acknowledgments}
We thank A. Prabhu for collaboration during the early stages of this work. We thank D. Dunsky, N. Gendler, S. Gori, S. Heeba, S. Hong, A. Parikh, N. Rodd, B. Safdi, G. Villadoro, and N. Weiner for useful discussions. We thank S. Gori, M. Kongsore, and A. Prabhu for comments on a draft of the manuscript\journal{, and S. Stelzl for comments on Fig.~\ref{fig:max_N}}. JTR is supported by the NSF award
PHY-2210498 and the BSF grant 2024091. 
This work was initiated in part at the Aspen Center for Physics, which is supported by National Science Foundation (NSF) grant PHY-2210452.
SK is supported in part by the NSF grant PHY-2310572.
SK acknowledges the hospitality of the Kavli Institute for Theoretical Physics (KITP) while the work was in progress.
This research was supported in part by the NSF grant PHY-2309135 to KITP\@. The Center for Computational Astrophysics at the Flatiron Institute is supported by the Simons Foundation.

\appendix
\section{General Considerations of Axiverse EFT}\label{sec:general}
The Lagrangian containing $\N$ axions $\theta_{i=1,\cdots,\N}$ is given by (see, {\it e.g.},~\cite{Mehta:2021pwf, Gendler:2023kjt}),
\es{eq:axiverse_general}{
{\cal L}_{\rm Axiverse} \supset {1\over 2} \sum_{i,j}\mpl^2 \partial_\mu\theta_i K_{ij} \partial^\mu \theta_j - \sum_{I,j} \Lambda_I^4 \left(1-\cos(\bar{Q}_{Ij}\theta_j +\delta_I)\right) \\+ \sum_{A,i} \tilde{c}_{A,i} {\cal O}^{(5)}_A \theta_i + \sum_{A,i,j} \tilde{d}_{A,ij}\partial^\mu\theta_i \partial^\nu \theta_j {\cal O}^{(6)}_{A,\mu\nu},
}
where the indices $A$, $I$, $\{i,j\}$ run over the SM operators, instantons generating axion potentials, and the different axions, respectively.
${\cal O}^{(5)}_A$ and ${\cal O}^{(6)}_A$ are the various SM operators that determine dimension-5 and dimension-6 axion-SM couplings, respectively.
For example, for the gluon coupling ${\cal O}^{(5)}_G = \alpha_s G\tilde{G}/(8\pi) $, for the fermion coupling, ${\cal O}^{(5)}_\Psi = \partial_\mu(\bar{\Psi}\gamma^\mu\Psi)$ (where we have suppressed the flavor indices), and for the dimension-6 coupling to the Higgs ${\cal O}^{(6)}_{A,\mu\nu} = |H|^2g_{\mu\nu}$.
For each SM operator $A$, $\tilde{c}_{A}$ and $\tilde{d}_A$ are $\N$-vector and $\N\times \N$ matrix, respectively.
Both the instanton charge matrix $\bar{Q}_{Ij}$ and anomalous gauge boson couplings consist of integers such that ${\cal L}_{\rm Axiverse}$ enjoys discrete shift symmetries $\theta_i \rightarrow \theta_i + 2\pi$.
For concreteness, we have modeled the periodicity of the axion potential by a cosine.
This axion potential can come from some UV physics, such as confining dark gauge groups or charged particles in extra dimensional theories~\cite{Arkani-Hamed:2003xts}.
Furthermore, in a given model it could be some other periodic function $f(\bar{Q}_{Ij}\theta_j +\delta_I)$, and the next discussion does not rely on the precise nature of this function.
We will focus on the case where the number of instantons is larger than the number of axions $\N$, as seen in some string theory constructions~\cite{Mehta:2021pwf}.

It is possible to write ${\cal L}_{\rm Axiverse}$ in a more convenient form, under the assumption that the sizes of the instanton-generated potentials are hierarchical, $\epsilon_I \equiv \Lambda_{I+1}^4/\Lambda_I^4 \ll 1$~\cite{Gendler:2023kjt}.
With this assumption, we can select $\N$ leading instantons with a charge matrix $Q_{ab}$, obtained by removing any row from $\bar{Q}_{Ia}$ that is a linear combination of the rows above it.
We can also use $2\pi$-shifts of $\theta_i$ to remove a certain number of arbitrary phases $\delta_I$.
With this procedure, the axion potential from ${\cal L}_{\rm Axiverse}$ can be written as~\cite{Gendler:2023kjt},
\es{}{
V(\theta)= - \sum_{i,j=1}^{\N} \Lambda_i^4 \left(1-\cos(Q_{ij}\theta_j)\right) + \cdots
}
where $\cdots$ contains subleading instanton contributions that potentially also contain CP-violating contributions.
We will ignore these subleading contributions below.
The reduced charge matrix $Q_{ij}$ is full rank by construction.
Therefore, one can perform an orthogonal basis rotation (Schur decomposition), $\phi = P \theta$ to diagonalize the kinetic matrix: $(P^{-1})^T K P^{-1} = D$, with $D$ some diagonal matrix, and put the charge matrix $Q$ in a lower triangular form.
One can redefine $\phi \rightarrow D^{-1/2}\phi/\mpl$ to make the kinetic term canonically normalized. 
Defining $QP^{-1}D^{-1/2} \equiv q$, $\tilde{c}_A (P^{-1})D^{-1/2} \equiv c_A$, and $D^{-1/2}(P^{-1})^T \tilde{d}_A (P^{-1})D^{-1/2} \equiv d_A$, we write~\eqref{eq:axiverse_general} as
\es{}{
{\cal L}_{\rm Axiverse} \supset {1\over 2} \sum_i\partial_\mu\phi_i  \partial^\mu \phi_i - \sum_{i,j=1}^{\N} \Lambda_i^4 \left(1-\cos\left(q_{ij}{\phi_j\over \mpl}\right)\right) \\+ \sum_{A,i} {c_{A,i} \over \mpl} {\cal O}^{(5)}_A \phi_i + \sum_{A,i,j} d_{A,ij}\partial^\mu\theta_i \partial^\nu \theta_j {\cal O}^{(6)}_{A,\mu\nu},
}
where $q_{ij}$ is lower-triangular and is not necessarily integer-valued.
Also, the coefficient matrix $d$ is not diagonal in general.
To the leading order in $\epsilon_I$, the mass matrix for axions $\phi_i$ is diagonal with the mass $m_i = \Lambda_i^2 q_{ii}/\mpl$.
Therefore, for subdominant axion self-interactions, the above can be simplified further
\es{eq:axiverse_simplified}{
{\cal L}_{\rm Axiverse} \supset {1\over 2} \sum_i\partial_\mu\phi_i  \partial^\mu \phi_i - {1\over 2}\sum_i m_i^2 \phi_i^2 + \sum_{A,i} {c_{A,i} \over \mpl} {\cal O}^{(5)}_A \phi_i + \sum_{A,i,j} d_{A,ij}\partial^\mu\theta_i \partial^\nu \theta_j {\cal O}^{(6)}_{A,\mu\nu},
}

To compute the cosmological abundance, it is more convenient to perform a final change of basis.
Suppose the axions couple to $\N_{\rm SM}$ (assumed to be $\leq \N$) SM dimension-5 operators, ordered as $\{ {\cal O}_1^{(5)},\cdots, {\cal O}_{\N_{\rm SM}}^{(5)}\}$.
In this ordering, close to the reheat temperature $\trh$, where freeze-in is most efficient, the axion freeze-in production rate via processes mediated by ${\cal O}_i^{(5)}$ is larger or equal to that mediated by ${\cal O}_j^{(5)}$ for $i\leq j$.
Each of $c_{1},\cdots, c_{\N_{\rm SM}}$ can be thought of as vectors in $\N$-dimensional axion vector space. 
Suppose among these $\N_{\rm SM}$ vectors, $\N_{\rm ind}\leq \N_{\rm SM}$ are linearly independent.
To construct the new basis, we perform the following steps, similar to a Gram-Schmidt process.
To begin, we define a new axion state $\bar{\phi}_1$ via
\es{}{
\sum_i c_{1,i}\phi_i = \bar{\phi}_1\left({\sum_i c_{1,i}^2}\right)^{1/2}.
}
We also relabel $c_1$ as $\bar{c}_1$.
Next, if $c_2$ is linearly independent from $\bar{c}_1$, we express it as a linear combination of $\bar{c}_{1}$ and $\bar{c}_{2}$ where $\bar{c}_2\cdot \bar{c}_1 = 0$.
The vector $\bar{c}_2$ defines another axion state $\bar{\phi}_2$ in analogy with the above equation.
Thus, both $\bar{\phi}_1$ and $\bar{\phi}_2$ couple to the operator ${\cal O}_2^{(5)}$.
On the other hand, if $c_2$ is proportional to $\bar{c}_1$, it does not lead to a new axion state.
Rather, it just describes the coupling of $\bar{\phi}_1$ with ${\cal O}_2^{(5)}$.

Repeating these steps gives a basis of axions $\{\bar{\phi}_1,\bar{\phi}_2,\cdots, \bar{\phi}_{\N_{\rm ind}}\}$ that couple to the SM operators, while the other ${\cal N}-{\cal N}_{\rm ind}$ axions do not.
In this rotated basis, $\bar{\phi}_1$, in general, couples to all of ${\cal O}_{i=1,\cdots,\N_{\rm SM}}^{(5)}$.
On the other hand, $\bar{\phi}_2$ couples to only ${\cal O}_{i=m,\cdots,\N_{\rm SM}}^{(5)}$, where $m$ is such that all $c_{i<m}$ are linearly dependent on $c_1$.
Similarly, $\bar{\phi}_3$ can be defined.
Given the operator ordering $\{ {\cal O}_1^{(5)},\cdots, {\cal O}_{\N_{\rm SM}}^{(5)}\}$ discussed above, a lower bound on the yield of freeze-in $\bar{\phi}_1$ production can be estimated by considering ${\cal O}_1^{(5)}$ only, while that of $\bar{\phi}_2$ production can be estimated by considering ${\cal O}_m^{(5)}$ only, and so on.

Having performed the basis rotation at dimension five, we cannot, in general, further reduce the dimension-6 couplings.
Hence, we treat each $d_{A,ij}$ as a rank $\N$ matrix and compute the axion abundance using that.
Given $\N$ can be much larger than $\N_{\rm ind}$, the dimension-6 operators can produce a parametrically larger abundance of axions compared to dimension-5 operators.

In the above, we assumed $\N\geq \N_{\rm SM}$. 
However, for $\N< \N_{\rm SM}$, one can follow similar steps.
If $\N\leq \N_{\rm ind}$, there would be no sterile axions.
Otherwise, there will be $\N- \N_{\rm ind}$ sterile axions as before.

\subsection{An Explicit Example}\label{sec:example}
Suppose we have three axions coupled to $SU(3)_c$ and $U(1)_Y$ via,
\es{eq:toy}{
\mathcal{L} \supset {1\over 2}(\partial\phi_1)^2 + {1\over 2}(\partial\phi_2)^2 + {1\over 2}(\partial\phi_3)^2 +  \dfrac{\alpha_s}{8\pi f} (\phi_1+\phi_2+\phi_3) \GG  +  \dfrac{\alpha_1}{8\pi f} (\phi_1+2\phi_2+\phi_3)\BB.
}
Following the above discussion ${\cal N}=3$ with ${\cal N}_{\rm SM}={\cal N}_{\rm ind}=2$.
The gluon coupling $c_1 = (1,1,1)\equiv \bar{c}_1$ determines a new state
\es{}{
\bar{\phi}_1 = {1\over \sqrt{3}}(\phi_1+\phi_2+\phi_3).
}
On the other hand, the $U(1)_Y$ coupling defines,
\es{}{
\bar{\phi}_2 = {1\over \sqrt{6}}(-\phi_1+2\phi_2-\phi_3),
}
with
$c_2 = (4/3)\times (1,1,1)+(1/3)\times (-1,2,-1)$ expressed in terms of $\bar{c}_1$ and $\bar{c}_2\equiv (-1,2,-1)$.
Finally, the state
\es{}{
\bar{\phi}_3 = {1\over\sqrt{2}}(-\phi_1+\phi_3)
}
does not couple to either $SU(3)_c$ or $U(1)_Y$.
Therefore, the Lagrangian in~\eqref{eq:toy} in the rotated basis is given by
\es{}{
{\cal L} \supset {1\over 2}(\partial\bar{\phi}_1)^2 + {1\over 2}(\partial\bar{\phi}_2)^2 + {1\over 2}(\partial\bar{\phi}_3)^2 + \dfrac{\sqrt{3}\alpha_s}{8\pi f}  \bar{\phi}_1\GG + \dfrac{\alpha_1}{8\pi f}\left({4\over \sqrt{3}}\bar{\phi}_1 + \sqrt{{2\over 3}}\bar{\phi}_2\right)\BB.
}
Note, since the $3\times 3$ matrix rotating $\phi$ basis into $\bar{\phi}$ basis is orthogonal, the kinetic term in the new basis remains diagonal.
Since for the temperatures of interest the production from $SU(3)_c$ interaction dominates over $U(1)_Y$, to a good approximation, we can ignore the interaction of $\bar{\phi}_1$ with $U(1)_Y$.
Thus, to compute a conservative lower limit on the axion abundance, we take $\bar{\phi}_1$ and $\bar{\phi}_2$ to be coupled to only $SU(3)_c$ and $U(1)_Y$, respectively.

\section{Comparing Derivative and Pseudoscalar Basis}
\label{app:deriv-pseudo}
Consider the Lagrangian in the derivative basis,
\es{eq:toy_lag}{
{\cal L} \supset {1\over 2f_a} \partial_\mu a \bar{\psi}\gamma^\mu \gamma_5 \psi + i\bar{\psi}\gamma^\mu \partial_\mu \psi + (y h - m) \bar{\psi}\psi.
}
Here we have added a bare fermion mass.
We can remove the derivative on the axion in two ways: (1) by using EOM and IBP or (2) by using field redefinition.
\subsection{IBP and EOM}
The EOM at LO without involving the axion interaction is sufficient for this purpose.
The fermion EOMs are:
\es{}{
i\gamma^\mu \partial_\mu \psi + (yh-m)\psi = 0,\\
-i \partial_\mu \bar{\psi}\gamma^\mu + (yh-m)\bar{\psi}=0.
}
We can use IBP to write~\eqref{eq:toy_lag} as (after dropping a boundary term),
\es{}{
{\cal L}\supset -{1\over 2f_a}a \partial_\mu(\bar{\psi}\gamma^\mu \gamma_5\psi) + i\bar{\psi}\gamma^\mu \partial_\mu \psi + (y h - m) \bar{\psi}\psi.
}
Using the EOMs this can be written as,
\es{eq:pseudo}{
{\cal L}\supset {i a\over f_a}(yh-m)\bar{\psi}\gamma_5 \psi + i\bar{\psi}\gamma^\mu \partial_\mu \psi + (y h - m) \bar{\psi}\psi.
}
\subsection{Field Redefinition}
We can perform a rotation,
\es{}{
\psi \rightarrow \exp(i\alpha(x)\gamma_5)\psi,~~\bar{\psi} \rightarrow \bar{\psi} \exp(i\alpha(x)\gamma_5).
}
This implies
\es{}{
i\bar{\psi}\gamma^\mu \partial_\mu \psi \rightarrow i\bar{\psi}\gamma^\mu \partial_\mu \psi - \bar{\psi}\partial_\mu \alpha \gamma^\mu \gamma_5 \psi,\\
(yh-m)\bar{\psi}\psi \rightarrow (yh-m)\bar{\psi}\exp(2i\alpha(x)\gamma_5)\psi.
}
Choosing $\alpha = a/(2f_a)$, removes the derivative coupling and introduces the pseudoscalar coupling.
Therefore in this pseudoscalar basis the Lagrangian is given by,
\es{}{
{\cal L}\supset i\bar{\psi}\gamma^\mu\partial_\mu\psi + {ia\over f_a}(yh-m)\bar{\psi}\gamma_5\psi + (yh-m)\bar{\psi}\psi.
}
This matches with the IBP+EOM manipulation.
\subsection{Comparison of Cross Sections}
We should get the same answer both in the pseudoscalar and derivative basis for axion freeze-in production.
To check that, note~\eqref{eq:pseudo} implies there are three diagrams: one contact diagram with a quartic vertex and two other with the topology shown in Fig.~\ref{fig:diag}.
The matrix elements are given by (with appropriate notations for particle momenta $p_{1}$ to $p_{4}$),
\es{}{
i{\cal M}_{\rm con} = (i) (i y/f_a) \bar{v}(p_1)\gamma_5 u(p_2),
}
\es{}{
i{\cal M}_{1} = (i y) (m/f_a) \bar{v}(p_1){i\over \slashed{p}_2-\slashed{p}_3-m}\gamma_5 u(p_2),
}
\es{}{
i{\cal M}_{2} = (i y) (m/f_a) \bar{v}(p_1)\gamma_5{i\over \slashed{p}_2-\slashed{p}_4-m}u(p_2).
}
The cross section following from these three amplitudes is given by,
\es{}{
\sigma = {y^2 \over 32 \pi f_a^2} + {y^2 m^2 \over 16f_a^2\pi s}\ln(m^2/s).
}

We can also evaluate this directly in the derivative basis.
The amplitudes are given by,
\es{}{
i{\cal M}_{3} = (iy) \bar{v}(p_1) {i \over \slashed{p}_4 - \slashed{p}_1 - m_i}\gamma^\nu  \gamma^5 u(p_2) {i\over 2f_a}(i p_{3\nu}),
}
and
\es{}{
i{\cal M}_{4} = (iy) \bar{v}(p_1)\gamma^\nu  \gamma^5 {i \over \slashed{p}_3 - \slashed{p}_1 - m_j} u(p_2) {i\over 2f_a}(i p_{3\nu}).
}
The cross section is thus given by
\es{}{
\sigma = {y^2 \over 32 \pi f_a^2} + {y^2 m^2 \over 16 \pi f_a^2 s}\ln(m^2/s),
}
matching with the result obtained using the pseudoscalar basis.

\bibliographystyle{JHEP}
\bibliography{references}
\end{document}